\documentclass{sigplanconf}

\usepackage{tikz}
\usetikzlibrary{calc,positioning}
\usepackage{graphicx}
\usepackage{amsmath}
\usepackage{amssymb}
\usepackage{cmll}
\usepackage{amsthm}
\usepackage{stmaryrd}
\usepackage{MnSymbol}

\newcommand{\ve}[1]{*+[o][F]{#1}}

\newcommand{\stcomp}[3]{\exists #1.\, [\, #2 \mathrel{
\vvbar} #3\, ]}

\newcommand{\Eta}{{\rm H}}
\newcommand{\Ga}{\Gamma}
 
  \newcommand{\Equiv}{{\rm Equiv}}
 \newcommand{\Prob}{{\rm Prob}}
\newcommand{\drp}[2]{d_v[{#1}; #2]}

\newcommand{\eeq}{\equiv}
\newcommand{\ese}{\rm ese}
\newcommand{\Ese}{\rm Ese}
\newcommand{\er}{\it er}
\newcommand{\edc}{\rm edc}
\newcommand{\ef}{\rm ef}
\newcommand{\ESE}{{\PES}_\equiv}

\newcommand{\FAME}{{\cal F}\!am_\equiv}
\newcommand{\SFAME}{{\cal S\!F}\!am_\equiv}
\newcommand{\GES}{{\cal G}}
\newcommand{\EDC}{{\cal E\!D\!C}}

\newcommand{\ges}{{\it ges}}

\newcommand{\tri}{\unlhd}
\newcommand{\A}{{\mathcal A}}
\newcommand{\B}{{\mathcal B}}
\newcommand{\C}{{\mathcal C}}
\newcommand{\D}{{\mathcal D}}

\newcommand{\sncirc}{\oast} 

\newtheorem{theorem}{Theorem}[section]
\newtheorem{corollary}[theorem]{Corollary}
\newtheorem{prop}[theorem]{Proposition}

\newtheorem{example}[theorem]{Example}

\theoremstyle{remark}

\newtheorem*{theorem*}{Theorem}

\def\endex{{\hspace*{\fill}\hbox{$\Box$}}}

 \usepackage[all]{xy}

\newcommand{\emptygame}{\emptyset}

\newcommand{\opmove}{\ominus}
\newcommand{\plmove}{\oplus}

\def\profto{\!\!\!\xymatrix@C-.75pc{\ar[r]|-{\! +\!} &}\!\!\! }

\def\all{\forall}

\newdir{pb}{:(1,-1)@^{|-}}
\def\pb#1{\save[]+<16 pt,0 pt>:a(#1)\ar@{pb{}}[]\restore}

\newcommand{\co}{\mathbin{{\it co}}}
\newcommand{\vvbar}{{\mathbin{\parallel}}}
\newcommand{\scirc}{{{\odot}}}
\newcommand{\cov}{{{\mathrel-\joinrel\subset}}}

\newcommand{\longcov}[1]{{\stackrel{#1}{\mathrel-\joinrel\relbar\joinrel\subset\,}}}

\renewcommand{\max}{\it top}

\newcommand{\imc}{\rightarrowtriangle}
\newcommand{\setdif}{\setminus}

\newcommand{\sig}{\sigma}

\newcommand{\fsubseteq}{\subseteq_{\rm fin}}

\newcommand{\CC}{{\rm C\!\!C}}
\newcommand{\cc}{\gamma}
\newcommand{\pol}{{\it pol}}

\newcommand{\F}{{\mathcal F}}
\newcommand{\G}{{\mathcal G}}

\newcommand{\bel}{\sqsubseteq}
  \newcommand{\spanpl}[5]{{\xymatrix{
    & {#3}\ar@{_{(}->}[dl]_{#2}\ar[dr]^-{#4} &\\
      {#1} && {#5}
  }}}

\newcommand{\PES}{{\cal E}}

\def\Con{{\rm Con}}

\newcommand{\Fam}{{\cal F}}

\newcommand{\conf}[1]{\:\!{\cal C}(#1)}
\newcommand{\iconf}[1]{\:\!{\cal C}^\infty(#1)}

\newcommand{\parrow}{\rightharpoonup}

\newcommand{\arr}[1]{{{\stackrel{#1}{\longrightarrow}}}}
\newcommand{\id}{{\rm id}}

\newcommand{\set}[2]{{\{  #1\  | \  #2 \} }}
\newcommand{\setof}[1]{{\{ #1 \} }}

\newcommand{\smallcat}[1]{\mathbb{#1}}

\newcommand{\eqdef}{\mathrel{=_{\mathrm{def}}}}

\newcommand{\iso}{\cong}

\renewcommand{\P}{\smallcat{P}}

\def\mxxth{\mathsurround=0pt}
\dimendef\dimenxx=0
\def\openup{\afterassignment\xxpenup\dimenxx=}
\def\xxpenup{\advance\lineskip\dimenxx
  \advance\baselineskip\dimenxx \advance\lineskiplimit\dimenxx}
\def\eqalign#1{\,\vcenter{\openup1\jot \mxxth
  \ialign{\strut\hfil$\displaystyle{##}$&$\displaystyle{{}##}$\hfil
     \crcr#1\crcr}}\,}

\newif\ifdtxxp
\def\displxxy{\global\dtxxptrue \openup1\jot \mxxth
  \everycr{\noalign{\ifdtxxp \global\dtxxpfalse
      \vskip-\lineskiplimit \vskip\normallineskiplimit
      \else \penalty\interdisplaylinepenalty \fi}}}
\def\displaylines#1{\displxxy
  \halign{\hbox to\displaywidth{$\hfil\displaystyle##\hfil$}\crcr
      #1\crcr}}

\newskip\mycntring \mycntring=0pt plus 1000pt minus 1000pt

\def\leqalignno#1{\displxxy \tabskip=\mycntring
  \halign to\displaywidth{\hfil$\displaystyle{##}$\tabskip=0pt
      &$\displaystyle{{}##}$\hfil\tabskip=\mycntring
      &\kern-\displaywidth\rlap{$##$}\tabskip=\displaywidth\crcr
      #1\crcr}}

\newcommand{\ie}{{\it i.e.}}
\newcommand{\eg}{{\it e.g.}}
\newcommand{\cf}{{\it cf.}}

\newcommand{\viz}{{\it viz.}}

\newdir{C}{%
!/4.5pt/@{( }*:(1,-.2)@{}*:(1,+.2)@_{}}

\newdir{|>}{%
!/4.5pt/@{|}*:(1,-.2)@{>}*:(1,+.2)@_{>}}

\begin{document}

\setlength{\pdfpageheight}{\paperheight}
\setlength{\pdfpagewidth}{\paperwidth}

\conferenceinfo{Logic in Computer Science 2016}{New York, USA}
\copyrightyear{2016} 
\copyrightdata{
} 
\doi{%2603088.2603141
}





\title{Strategies with Parallel Causes}

\authorinfo{Marc de Visme}
           {Ecole Normale Sup\'erieure de 
Paris, France}
           {}
\authorinfo{Glynn Winskel}
	   {Computer Laboratory, University of Cambridge, UK}
           {}
 
\maketitle

\begin{abstract}
In a distributed game we imagine a team Player engaging a team Opponent in a distributed fashion.  Such games and their strategies have been formalised in concurrent games based on event structures.  However 
there are limitations in founding strategies on traditional event structures. Sometimes a probabilistic distributed strategy relies on certain benign races where, intuitively, several members of team Player may race each other to make a common move.  Although there are event structures which support such parallel causes, in which an event is enabled in several compatible ways, they do not support an operation of hiding central to the composition of strategies; nor do they support probability adequately.  An extension of traditional event structures is devised which supports parallel causes and hiding, as well as the mix of probability and nondeterminism needed to account for probabilistic distributed strategies.  The extension is tested in the construction of a bicategory of probabilistic distributed strategies with parallel causes.  The bicategory is rich in operations relevant to probabilistic as well as deterministic parallel programming. 
\end{abstract}

\section{Introduction}
This article considers probabilistic distributed games between two teams, Player and Opponent. 
To set the scene, 
imagine  a simple distributed game in which team  Opponent can perform two moves, called 1 and 2, far apart from each other, and that team Player can just make one move, 3.  Suppose that for Player to win they must make their move iff Opponent makes one or more of their moves.  Informally Player can win by assigning two members of their team, one to watch out for the Opponent move 1 and  the other Opponent move 2.  When either watcher sees their respective Opponent move they run back and make the Player move 3.  Opponent could possibly play both 1 and 2 in which case both watchers would run back and could make their move together.  Provided the watchers are perfectly reliable this provides a winning strategy for Player.  No matter how Opponent chooses to play or not play their moves, Player will win;  if Opponent is completely inactive the watchers wait forever but then Player does win, eventually.  

We can  imagine variations in which the watchers are only reliable with certain probabilities  
with a consequent reduction in the probability of Player winning against Opponent strategies. 
In such a probabilistic 
 strategy Player can only determine probabilities of their moves conditionally on those of Opponent. Because Player has no say in the 
 probabilities of Opponent moves beyond those determined by causal dependencies of the strategy we are led to a {\em Limited 
Markov Condition}, of the kind discussed in~\cite{Pearl}:
\begin{quote}
{\bf (LMC)}
In a situation $x$ in which both a Player move $\plmove$ and an Opponent move $\opmove$  could occur individually,
if the Player move and the Opponent move are causally independent, then they are probabilistically independent; in a strategy for Player, 
$\Prob(\plmove \mid x,  \opmove) = \Prob(\plmove \mid  x)$.  
\end{quote}
The LMC is borne out in the game of ``matching pennies'' where Player and Opponent in isolation, so independent from each other, each make their choice of head or tails.  Note we do not expect that  in all strategies for Player that 
two causally independent Player moves are necessarily probabilistically independent;  in fact, looking ahead, because composition of strategies involves hiding internal moves such a property would not generally be preserved by composition.  
 
Let us try to describe the informal strategy above in terms of event structures. 
 In `prime' event structures in which causally dependency is expressed a partial order, an event is causally dependent on a unique set of events, \viz~those events below it in the partial order.    For this reason within prime event structures we are forced to split the Player move into two events one for each watcher making the move, one $w1$ dependent on Opponent move 1 and the other $w2$ on Opponent move 2.  The two moves of the two watchers stand for  the same move in the game.  Because of this they are in conflict (or inconsistent) with each other.
 We end up with the event structure drawn below:
 $$
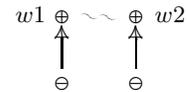
\small
w1\xymatrix@R=14pt@C=5pt{
\plmove
\ar@{~}[rr]&&\plmove
\\
\ar@{|>}[u] \opmove
& &\ar@{|>}[u]\opmove
}\,w2
$$
The polarities + and $-$ signify moves of Player and Opponent, respectively.  
The arrows represent the (immediate) causal dependencies and the wiggly line conflict. As far as purely nondeterministic behaviour goes, we have expressed the informal strategy reasonably well:   no matter how Opponent makes or doesn't make their moves any maximal play of Player is assured to win.  However
consider assigning conditional probabilities to the watcher moves.  Suppose the probability of $w1$ conditional on 1 is $p_1$, \ie~$\Prob(w1 \mid  1) = \Prob(w1,1\mid  1) = p_1$ and that similarly for $w1$ its conditional probability $\Prob(w2\mid 2)=p_2$.
Given that move $w1$ of Player  and move $2$ of Opponent are causally independent, from (LMC) we expect that $w1$ is 
 probabilistically independent of move $2$, \ie
$$
\Prob(w1\mid 1,2)=\Prob(w1\mid 1)=p_1\,;
$$
 whether Opponent chooses to make move 2 or not should have no influence on the watcher of move 1.
Similarly,
$$
\Prob(w2\mid 1,2)=\Prob(w2\mid 2)=p_2 \,.
$$
But $w1$ and $w2$ are in conflict, so mutually exclusive, and can each occur individually when 1 and 2 have occurred ensuring that 
$$
p_1+p_2 \leq 1
$$
---we haven't  insisted on one or the other occurring, the reason why we have not written equality.
The best Player can do is  
assign $p_1=p_2 = 1/2$. 
Against a counter-strategy with Opponent playing one of their two moves with probability  1/2 this strategy 
only wins half the time.   We have clearly failed to express the informal winning strategy accurately!  
 
 Present notions of ``concurrent strategies,'' the most general of which are presented in~\cite{Probstrats}, are or can be expressed using prime event structures. 
If we are to be able to express the intuitive strategy which wins with certainty
we need to develop distributed probabilistic strategies which allow  {\em parallel causes}
in which an event can be enabled in distinct but compatible ways.  
 `General' event structures  are one such model
~\cite{evstrs}.  In the informal strategy described in the previous section both Opponent  moves would individually enable the Player move, with all events being consistent, illustrated below:
 $$
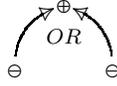
\small
\xymatrix@R=1pt@C=3pt{
 &\plmove&
\\
&{OR}&\\
\ar@{|>}@/^0.5pc/[uur] \opmove
& &\ar@{|>}@/_0.5pc/[uul]\opmove
}
$$
 But as we shall see general event structures do not support an appropriate operation of hiding central to the composition of strategies.  Nor is it clear how within general event structures one could express the variant of the strategy above in which the two watchers succeed in reporting 
 with different probabilities. 
 
It has been necessary to  
develop a new model---{\em event structures with disjunctive causes} (edc's)---which support hiding and probability adequately, and into which both prime and general event structures embed.  The new model provides a foundation on which 
 to build a  theory and rich language of probabilistic distributed strategies with parallel causes.  Without probability, 
 it provides a new bicategory of 
 deterministic parallel strategies which includes, for example, a deterministic strategy for computing ``parallel or"---Section~\ref{sec:edcstrats}.
 
{\it Full proofs can be found in~\cite{ecsym-notes}. Appendix~A summarises the simple instances of concepts we borrow from enriched categories~\cite{kelly} and 2-categories~\cite{Power2-cats}.}
  


\section{Event structures}

Event structures describe a process, or system, in terms of its possible event occurrences, their causal dependencies and consistency. The simplest form, `prime' event structures, are a concurrent, or distributed, analogue of trees; 
though in such an event structure the individual `branches' are no longer necessarily sequences but have the shape of a partial order of events.

\subsection{Prime event structures} 
A {\em (prime) event structure}  comprises $(E, \leq, \Con)$,   consisting of a set $E$ of {\em events} (really event occurrences) which are
partially ordered by $\leq$, the {\em causal dependency
relation},
and a  nonempty {\em consistency relation} $\Con$ consisting of finite subsets of $E$.  The relation
$e'\leq e$ expresses that event $e$ causally depends on the previous occurrence of event $e'$.  That a finite subset of events is consistent conveys that its events can occur together by some stage in the evolution of the process.  
Together the relations satisfy several  axioms:
$$
\eqalign{
&[e]\eqdef \set{e'}{e'\leq e}\hbox{ is finite for all } e\in E,\cr
&\setof{e}\in\Con \hbox{  for all } e\in E,\cr
&Y\subseteq X\in\Con \hbox{ implies }Y\in \Con,\ \hbox{ and}\cr
&X\in\Con \ \&\  e\leq e'\in X \hbox{ implies } 
X\cup\setof{e}\in\Con.\cr}
$$
Given this understanding of an event structure, there is an accompanying notion of state, or history, those events that may occur  up to some stage in the behaviour of the process described.  A {\em configuration} is a, possibly infinite, set of events $x\subseteq E$ which is 
\begin{itemize} 
\item[]
{\em consistent:} $X\subseteq x  \hbox{ and }  X \hbox{ is finite}  \hbox{ implies } X\in\Con$\,,
and
\item[]
{\em down-closed:}
$ e'\leq e\in x  \hbox{ implies } e' \in x$.
\end{itemize} 
A configuration inherits a partial order from the ambient event structure, and represents a possible  partial-order history.  

Two events $e, e'$ are considered to be causally independent, and called {\em concurrent} if  the set $\setof{e,e'}$ is in $\Con$ and neither event is causally dependent on the other.  
The relation of {\em immediate} dependency $e\imc e'$ means $e$ and $e'$ are distinct with $e\leq e'$ and no event in between. 
Write $\iconf E$ for the configurations of $E$ and $\conf E$ for its finite configurations.    For configurations $x,y$, we use   $x\cov y$ to mean $y$ covers $x$, \ie~
$x\subset y$   with nothing in between, and $x \longcov e y$ to mean $x\cup\setof e =y$ for an event $e\notin x$.
We sometimes use $x \longcov e$, expressing that event $e$ is enabled at configuration $x$, when $x \longcov e y$ for some $y$.  

It will be very useful to relate event structures by maps.  
 A {\em map} of event structures 
$f:E
\to E'$
 is a partial function  
$f$ from $E$ to $E'$ such that
the image of a configuration $x$ is a configuration $f x$ and any event of $f x$  arises as the image of a unique event of $x$.  Maps compose as partial functions. Write $\PES$ for the ensuing category. 

A map $f:E
\to E'$ reflects causal dependency locally, in the sense that if $e, e'$ are events in a configuration $x$ of $E$ for which $f(e')\leq f(e)$ in $E'$, then $e'\leq e$ also in $E$; the event structure $E$ inherits causal dependencies from the event structure $E'$ via the map $f$. Consequently,  a map 
preserves concurrency:  if two events are concurrent, then their images if defined are also concurrent.  In general a map of event structures need not preserve causal dependency; when it does and is total we say it is {\em rigid}. 
 
\subsection{General event structures}
A {\em general event structure}~\cite{thesis,evstrs} is a structure $(E, \Con, \vdash)$  where $E$ is a set of event occurrences, the consistency relation $\Con$ is a non-empty collection of finite subsets of $E$ satisfying
$$X\subseteq Y\in \Con \implies X\in\Con$$
and the {\em enabling relation} $\vdash\subseteq \Con\times E$ satisfies
$$
Y\in\Con\ \&\ Y\supseteq X \ \& \ X\vdash e \implies Y\vdash e\,.
$$ 
A {\em configuration} is a subset of $E$ which is 
\begin{itemize}
\item[]
 {\em consistent:}  $X\fsubseteq x \implies X\in\Con$ and 
\item[]
{\em secured:}\! $\all e\in x\exists  e_1, \cdots, e_n\in x.\ e_n=e\,\&
\all i\leq n. \setof{e_1, \cdots
, e_{i-1}}\vdash~e_i$.  
\end{itemize}
Again we write $\iconf E$ for the configurations of $E$ and $\conf E$ for its finite configurations.  

The
  notion of secured has been expressed through the existence of a securing chain to express an enabling of an event within a set which is a complete enabling in the sense that everything in the securing chain is itself enabled by earlier members of the chain.  One can imagine more refined ways in which to express complete enablings which are rather like proofs. 
 Later the idea that complete enablings are consistent partial orders of events in which all events are enabled by earlier events in the order---``causal realisations''---will play an important role in generalising general event structures to structures  supporting hiding and parallel causes. 
  
  A {\em map} $f:(E,\Con,\vdash)\to (E',\Con',\vdash')$ of general event structures  is a partial function $f:E\parrow E'$ such that 
  \begin{itemize}
\item[]
  $X\in\Con\implies fX \in \Con' \ \&\ \\
  \all e_1, e_2\in X.\ f(e_1)=f(e_2) \implies e_1=e_2$ 
  and 
  \item[]
  $X\vdash e \ \&\ f(e)\hbox{  is defined }\implies fX\vdash' f(e)$\,.
  \end{itemize}
 Maps compose as partial functions with identity maps being identity functions. Write $\GES$ for the category of general event structures.

We can characterise those families of configurations arising from a  
 general event structure. A {\em family of configurations} which comprises a  
family $\F$ of sets   such that
 \begin{itemize}
\item[]
if $X\subseteq\F$ is finitely compatible in $\F$ then  $\bigcup X\in F$; and
\item[]
if $e\in x\in \F$ then there exists a securing chain $e_1, \cdots, e_n=e$ in $x$ such that $\setof{e_1, \cdots, e_i} \in \F$ for all $i\leq n$.  
 \end{itemize}
 The latter condition is equivalent to saying (i) that whenever $e\in x\in \F$ there is a finite $x_0\in \F$  such that $e\in x_0\in \F$ and (ii) that if
$e, e'\in x$ and $e\neq e'$  then there is $y\in \F$ with $y\subseteq x$ s.t.~$e\in y \iff e'\neq y$.
The elements of the underlying set $\bigcup \F$  are its {\em events}.  
Such a 
 family is {\em stable} when for any compatible non-empty subset  $X$ of $\F$ its intersection $\bigcap X$ is a member of $\F$.   

A configuration $x\in\F$ is {\em irreducible} iff there is a necessarily unique $e\in x$ such that $\all y\in \F.\ e\in y \subseteq x$ implies $y=x$. 
Irreducibles coincide with complete join irreducibles w.r.t.~the order of inclusion. 
It is tempting to think of irreducibles as representing minimal complete enablings.  But, as sets, irreducibles both (1) lack sufficient structure:  
in the  formulation we are led to of minimal complete enabling as prime causal realisations,
 several prime realisations can have the same irreducible as their underlying set; 
and (2) are not general enough: there are prime realisations whose underlying set is not an irreducible.  

A map between families of configurations from $\F$ to $\G$ is a partial function $f:\bigcup\F \parrow \bigcup\G$ between their events such that  for any   $x\in\F$ its image $fx\in\G$   and  
  $$
 \all e_1, e_2\in x.\ f(e_1)=f(e_2) \implies e_1=e_2\,.
 $$
 Maps between general event structures satisfy this property.
 Maps of families compose as partial functions. 
 
The forgetful functor 
taking a general event structure to its family of configurations has a left adjoint, which constructs a canonical general event structure from a family:  given $\A$, a family of configurations with underlying events $A$,   construct a general event structure  
$
(A,\Con, \vdash)$  
with
\begin{itemize}
\item[]
$X\in \Con$ iff $X\fsubseteq y$, for some $y\in \A$, and
\item[]
$X\vdash a$ iff $a\in A$, $X\in\Con\ \&\ e\in y  \subseteq X\cup\setof a$, for some $y\in \A$.
\end{itemize}
  The above yields a coreflection of families of configurations in general event structures.  It 
cuts down to an equivalence between families of configurations and {\em replete} general event structures.
A general event structure $(E,\Con, \vdash)$ is {\em replete}   iff 
$$\eqalign{
&\all e\in E \exists X\in\Con.\ X\vdash e \,,\cr
&\all X\in\Con\exists x\in\conf E.\  X\subseteq x  \hbox{ and }\cr
&X\vdash e \implies \exists x\in \conf E.\ e\in x \ \&\ x\subseteq X\cup\setof e\,.}$$
The last condition is equivalent to stipulating that each minimal enabling $X\vdash e$, where $X$ is a minimal consistent set enabling $e$, corresponds to an irreducible configuration $X\cup\setof e$.

 \subsection{On relating prime and general event structures}\label{sec:relprimeandgen}
 Clearly a prime event structure $(P, \leq, \Con)$ can be identified with a (replete) general event structure $(P, \vdash, \Con)$ by  taking 
$$ 
 X\vdash p 
\hbox{ iff }
 X\in \Con \ \&\ [p]\subseteq X\cup\setof p\,.
 $$
Indeed under this identification  there is a full and faithful embedding of $\PES$ in $\GES$.  However (contrary to the claim in~\cite{evstrs}) there is no adjoint to this embedding.  This leaves open the issue of providing a canonical way to describe a general event structure as a prime event structure.  This issue has arisen as a central problem in reversible computation~\cite{ioana} and now more recently in the present limitation of   concurrent strategies described in the introduction.  A corollary of our work will be that the embedding of prime into general event structures does have a {\em pseudo} right adjoint, at the slight cost of enriching prime event structures with equivalence relations.  
\section{Problems  with general event structures}
Why not settle for general event structures as a foundation for distributed strategies?
Because they don't support hiding so composition of strategies; 
nor do they support probability 
generally enough.
\subsection{Probability and parallel causes}
We return to the general-event-structure description of the strategy in  the Introduction.  
To turn this into a probabilistic strategy for Player we should
 assign 
probabilities  to configurations 
conditional on 
Opponent moves.
%
The watcher of 1 is causally independent of Opponent move 2.  Given this we might expect that
the probability of the watcher of 1 making the Player move 3 should be probabilistically independent of move 2; after all, both moves 3 and 2 can occur concurrently from configuration $\setof 1$. Applying LMC naively would yield
$$
\Prob(1,3 \mid  1) = \Prob(1,2, 3 \mid  1,2)
\,.$$
But similarly, $\Prob(2,3\mid 2) = \Prob(1,2, 3 \mid  1,2)$, which forces $\Prob(1,3 \mid  1)  = \Prob(2,3\mid 2)$, \ie~that the conditional probabilities of the two watchers succeeding are the same!  In blurring the distinct ways in which move 3 can be caused we have obscured causal independence which has led us 
 to identify possibly distinct probabilities. 
%
\subsection{Hiding}
With one exception, all the operations  used in building strategies and, in particular, the bicategory of concurrent strategies~\cite{lics11} 
extend  to general event structures.  The one exception, that of hiding, is crucial in ensuring composition  of strategies yields  a bicategory.

Consider a general event structure with 
%
{\em events}   $a, b, c, d$ and $e$; 
%
{\em enabling}  (1) $ b,c \vdash e $               and          (2)  $d \vdash e$, with all events other than $e$ 
being enabled by the empty set; and 
%
{\em consistency} in which all subsets are consistent unless they contain the events $a$
and $b$ ---
the events $a$ and $b$ are in conflict.

Any configuration will satisfy the assertion
$$
 (a \wedge e) \implies d
 $$
because if $e$ has occurred it has to have been enabled by (1) or (2) and if
$a$ has occurred its conflict with $b$ has prevented the enabling (1), so $e$ can
only have occurred via enabling (2).

Now imagine the event $b$ is hidden, so allowed to occur invisibly in the
background. The configurations after hiding are those obtained by
hiding (\ie~removing) the invisible event $b$ from the configurations of the
original event structure. The assertion above will still hold of the
configurations after hiding.
There isn't a general event structure with events $a, c, d$ and $e$, and
configurations those which result when we hide (remove) $b$ from the
configurations of the original event structure. One way to see this is to
observe that amongst the configurations after hiding we have
$$
 \setof{c} \cov  \setof{c, e} \hbox{   and   }   \setof{c} \cov  \setof{a, c}
$$
where
 both  $ \setof{c, e}$  and  $ \setof{a, c}$ have upper bound  $ \setof{a, c, d, e}$, 
 and yet
 $\setof{ a, c, e}$ is not a configuration after hiding as it fails to satisfy the
assertion.
(In configurations of a general event structure if $x \cov y$ and $x \cov
z$ and $y $ and $z$ are bounded above, then $y \cup z$ is a configuration.)
Precisely the same problem can arise in the composition (with hiding) of strategies based on general event structures.


To obtain a bicategory of strategies with disjunctive causes we need to
support hiding. We need to look for structures more general
than general event structures. The example above gives a clue: the
inconsistency should be one of inconsistency between (minimal complete) 
enablings rather than
events.

\section{Adding disjunctive causes
}\label{sec:EDC}

To cope with disjunctive causes and hiding we must go beyond general event structures.
We  introduce structures in which we {\em objectify} cause
; a minimal complete 
enabling is no longer an instance of a relation but a structure that realises that instance (\cf~a judgement of theorem-hood in contrast to a  
proof).  This is in order to express inconsistency between minimal complete  enablings, inexpressible as inconsistencies on events,  that can arise when hiding.  

Fortunately we can do this while staying close to prime event structures.  The twist is to regard ``disjunctive events'' as comprising subsets of events of a prime event structure, the events  of which are now to be thought of as representing ``prime causes'' standing for 
minimal complete enablings.  Technically, we do this by extending prime event structures with an equivalence relation on  events.  

In detail, an {\em event structure with 
equivalence} (an \ese)
is a structure
$$(P, \leq, \Con, \eeq)$$
where $(P, \leq, \Con)$ satisfies the axioms of a (prime) event structure and $\eeq$ is an equivalence relation on $P$.

An \ese~dissociates the two roles of enabling and atomic action conflated in the events of a prime event structures. 
The intention is that the events $p$ of $P$, or really their corresponding down-closures $[p]$, describe minimal  complete enablings, {\em prime causes}, while the $\eeq$-equivalence classes of $P$ represent {\em disjunctive events}: $p$ is a prime cause of the disjunctive event $\setof p_\eeq$. Notice there may be several prime causes of the same event
and that these may be {\em parallel causes} in the sense that they are consistent with each other and  not  related in the order $\leq$.  

A {\em configuration} of the \ese~is a configuration of $(P, \leq, \Con)$ and we shall use the notation of earlier on event structures  $\iconf P$ and $\conf P$ for its configurations, respectively finite configurations.  However, we modify the relation of concurrency a little and say $p_1, p_2\in P$ are {\em concurrent} and write $p_1\!\!\co p_2$ iff $p_1\not\eeq p_2$ and $\setof{p_1, p_2}\in\Con$ and neither $p_1\leq p_2$ nor $p_2\leq p_1$.

When the equivalence relation $\eeq$ of an \ese~is the identity it is essentially a prime event structure. This view is reinforced in our choice of maps.
A map from \ese~$(P, 
\eeq_P)$ to $(Q, 
\eeq_Q)$ is a partial function $f:P\parrow Q$  which {\em preserves $\eeq$},  
 \ie
~if $p_1\eeq_P p_2$ then either both $f(p_1)$ and $f(p_2)$ are undefined or both defined with $f(p_1)\eeq_Q f(p_2)$), 
such that
for all $x\in\conf P$ 
\begin{enumerate}
\item[(i)]
the direct image $f x\in \conf Q$, and 
\item[(ii)]
$
\all p_1, p_2\in x. \ f(p_1)\eeq_Q  f(p_2) \implies p_1\eeq_P p_2\,.
$
\end{enumerate}
Maps compose as partial functions with the usual identities.  

%

It is not true that such maps preserve concurrency in general; they only do so locally w.r.t.~{\em unambiguous} configurations in which no two distinct elements are $\eeq$-equivalent.  

 We regard two maps $f_1,  f_2:P\to Q$ as equivalent, and write $f_1\equiv f_2$,  iff they are equi-defined and yield 
equivalent results, \ie

  if $ f_1(p)$ is defined then so is  $f_2(p)$ and $f_1(p) \eeq_Q f_2(p)$, and 

 if $f_2(p$) is defined then so is  $f_1(p)$ and $f_1(p) \eeq_Q f_2(p)$.
 
Composition respects $\equiv$:  if $f_1,  f_2:P\to Q$ with $f_1\equiv f_2$ and  $g_1,  g_2:Q\to R$ with $g_1\equiv g_2$, then $g_1f_1 \equiv g_2f_2$.  
Write $\ESE$ for the category of \ese's; it is {\em enriched} in the category of sets with equivalence relations---see Appendix~A. 

\Ese's support a hiding operation. 
Let $(P, \leq, \Con_P, \eeq)$ be an \ese.  Let $V\subseteq P$ be a $\eeq$-closed subset of `visible' events.
Define  
the {\em projection} of $P$ on $V$,   
to be 
$
P{\mathbin\downarrow} V\eqdef
(V, \leq_V, \Con_V, \eeq_V)
$, 
where
$v \leq_V v' \hbox{ iff } v\leq v' \ \&\ v,v'\in V$ and $X\in\Con_V \hbox{ iff }  X\in\Con\ \&\ X\subseteq V$
and
$v \eeq_V v' \hbox{ iff } v\eeq v' \ \&\ v,v'\in V$. 

 Hiding is associated with a factorisation of partial maps.
Let $$f: (P, \leq_P, \Con_P, \eeq_P)\to (Q, \leq_Q, \Con_Q, \eeq_Q)$$ be a  partial map between two \ese's. 
 Let 
$$V\eqdef \set{e\in E}{ f(e) \hbox{ is defined}}\,.$$
Then $f$  factors into the composition 
$$
\xymatrix{
P\ar[r]^{f_0}& P{\mathbin\downarrow} V \ar[r]^{f_1}& Q }
$$
of $f_0$, a partial map of \ese's taking $p\in P$ to itself if $p\in V$ and undefined otherwise, and $f_1$, a total map of \ese's acting like $f$ on $V$. We call $f_1$ the {\em defined part} of the partial map $f$.  Because $\eeq$-equivalent maps share the same domain of definition, $\eeq$-equivalent maps will determine the same projection and $\eeq$-equivalent defined parts. 
The factorisation is characterised to within isomorphism by the following 
  universal characterisation:  for any factorisation $
\xymatrix{
P\ar[r]^{g_0}& P_1 \ar[r]^{g_1}& Q }
$
where $g_0$ is partial and $g_1$ is total there is a  (necessarily total) unique map $h: P{\mathbin\downarrow} V\to P_1$  such that 
$$
\small
\xymatrix@R=12pt@C=20pt{
P\ar[r]^{f_0}\ar[dr]_{g_0}& P{\mathbin\downarrow} V\ar@{-->}[d]^h \ar[r]^{f_1}& Q \\
 & P_1\ar[ur]_{g_1}& 
}
$$
commutes. 
 
 The category $\ESE$ of \ese's supports hiding in the sense above.  We next 
 show how replete general event structures 
 embed 
  in \ese's
. 
 
\section{
A pseudo adjunction}

 The (pseudo) functor from $\GES$ to $\ESE$ 
 is quite subtle but arises as a right adjoint to a more obvious functor from $\ESE$ to $\GES$. 

Given an ese $(P,\leq, \Con,\eeq)$ we can construct a (replete) general event structure $\ges(P)\eqdef (E, \Con_E, \vdash)$ by taking 

$E = P_\eeq$, the equivalence classes under  $\eeq$;

$X\in \Con_E$ iff $\exists Y\in \Con. \  X= Y_\eeq$; and 

$X\vdash e$ iff $X\in \Con\ \&\ e\in E\ \&\ \\
\hbox{\ }  \quad \qquad \qquad \exists p\in P.\  e =\setof p_\eeq \ \&\ [p]_\eeq \subseteq X\cup\setof e$.

\noindent
The construction extends to a functor $\ges:\ESE\to \GES$ as maps between ese's preserve $\eeq$; the functor takes a map $f:P\to Q$ 
of ese's to the map $\ges(f):\ges(P)\to \ges(Q)$ obtained as the partial function induced on equivalence classes.  Less obvious is that there is a (pseudo) right adjoint to $\ges$.  
Its construction relies on 
extremal causal realisations which provide us with  
an appropriate notion of minimal complete enabling of events in a general event structure.
 
 \subsection{Causal realisations}
 Let $\A$ be a family of configurations with underlying set $A$.   
 
  A {\em (causal) realisation} of $\A$ 
comprises a partial order
$$
(E, \leq)\,,
$$
its {\em carrier}, such that the set $\set{e'\in E}{e'\leq e}$
is finite for all events $e\in E$, together with a function
$
\rho:E\to A
$
for which the
image
  $\rho x \in \A$ when $x$ is a down-closed subset of $E$.


A map between realisations $(E,\leq), \rho $ and $(E',\leq'),\rho'$  is a partial surjective function
$f:E\parrow E'$ which preserves down-closed subsets and satisfies
$\rho(e) = \rho'(f(e))$ when $f(e)$ is defined. 
 It   is convenient to write such a map as
$
\rho \succeq^f \rho'$.
Occasionally we shall write $\rho \succeq \rho'$, or the converse $\rho' \preceq \rho$, to mean there is a map of realisations from $\rho$ to  $\rho'$.
 
Such a map factors into a ``projection'' followed by a total map
$$\rho \succeq^{f_1}_1 \rho_0 \succeq^{f_2}_2 \rho'$$ where $\rho_0$ stands for the realisation $(E_0,\leq_0),\rho_0$ where
$$E_0 = \set{r\in R}{f(r) \hbox{ is defined}}\,,$$
the domain of definition of $f$, 
with $\leq_0$ the restriction of $\leq$, 
and $f_1$ is the inverse relation to the inclusion $E_0\subseteq E$,  and $f_2$ is the total function $f_2:E_0\to E'$. We are using $\succeq_1$ and $\succeq_2$ to signify the two kinds of maps. Notice that  $\succeq_1$-maps are reverse inclusions.  Notice too that $\succeq_2$-maps are exactly the total maps of realisations.  Total maps  $\rho \succeq_2^f \rho'$  are precisely those functions $f$ from the carrier of $\rho$ to the carrier of $\rho'$ which preserve down-closed subsets and satisfy $\rho = \rho' f$.

We shall say a  realisation $\rho$ is {\em extremal} when
$
\rho \succeq_2^f \rho'$ implies $f$ is an isomorphism,
for any realisation $\rho'$. 
 
In the special case where $\A$ is the family of configurations of a prime event structure, it is easy to show that an extremal realisation $\rho$ forms a bijection with a configuration of the event structure and that the order on the carrier coincides with causal dependency there.    

The construction is more interesting when $\A$ is the  family of configurations of a general event structure.  
In general, 
there is at most one map between extremal realisations.
 Hence extremal realisations of $A$ under $\preceq$ form a preorder.  The {\em order of extremal realisations} has as elements isomorphism classes of extremal realisations ordered according to the existence of a map between representatives of isomorphism classes.  As we shall see,
the order of extremal realisations 
forms a prime-algebraic domain~\cite{NPW} with complete primes represented by those extremal realisations which have a top element---a direct corollary of Proposition~\ref{prop:domianofrealzns} in the next section.  (We say a realisation has a top element when its carrier contains an element which dominates all other elements in the carrier.)


We provide examples illustrating the nature of extremal realisations.   In the examples it is convenient to describe families of configurations by general event structures,  taking advantage of the economic representation they provide.
\begin{example}{\rm
This and the following example shows that 
extremal realisations with a top do not correspond to irreducible configurations. 
Below, on the right we show a general event structure with irreducible configuration $\setof{a,b,c,d}$. On the left we show 
 two 
 extremals with tops $d_1$ and $d_2$ which both have the same irreducible configuration $\setof{a,b,c,d}$ as their image. The lettering indicates the functions associated with the realisations, \eg~events $d_1$ and $d_2$ in the partial orders map to $d$ in the general event structure.
$$\small
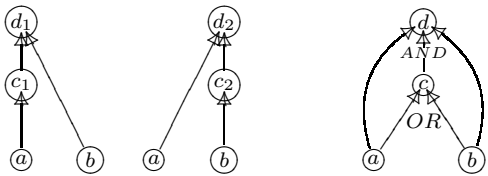
\xymatrix@R=3pt@C=4pt{
  \ve{d_1} 
  & {} & 
{} &{} & 
& {} 
& \ve{d_2}
{} &  {} & {} & {} & {} &
{} & \ve{d} & 
                \\
{} & {} & {} & {} & {} & {} &
        {} & {} & {} & {} &{} &
                {} & {} & {} \\ 
                \ve{c_1} \ar@{|>}[uu]^{
                } 
                & {} & 
&   {} &        
                & {} & \ve{c_2} \ar@{|>}[uu]_{
                }
& {}
& {} & {} & {} &
 {} & \ve{c} \ar@{|>}[uu]|{AND
} & {} 
                \\
{} & & {} &
        {} & & 
        {} & {} & {} & {} &{} &{} &
                {} & {OR}  & {} \\
 \ve{a} \ar@{|>}[uu] 
 & {} & \ve{b} 
 \ar@{|>}[uuuull]
&{} & \ve{a} 
\ar@{|>}[uuuurr] & {} & \ve{b} \ar@{|>}[uu] 
& {} & {} & {} & {} &
 \ve{a} \ar@{|>}[uur] \ar@{|>}@/^1pc/[uuuur] & {} & \ve{b} \ar@{|>}[uul] \ar@{|>}@/_1pc/[uuuul]        
        }$$
 }\end{example}
\begin{example}{\rm  On the other hand there are
 extremal realisations with top of which the image is   not   an irreducible configuration.  
Below the 
extremal with top on the left 
describes a situation where $d$ is enabled by $b$ and $c$ being enabled by $a$.
It has image the configuration $\setof{a,b,c,d}$ which is not irreducible, being the union of the two configurations $\setof a$ and $\setof{b,c,d}$. 
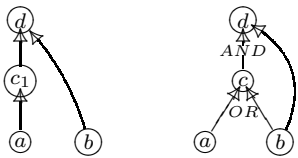
 $$
 \small\xymatrix@R=3pt@C=4pt{
                \ve{d} 
                & {} & 
                &
                        {} & {} & {} &
                                {} & \ve{d} & {} \\
\\
                \ve{c_1} 
                \ar@{|>}[uu]|{
                } & {} & 
                &
                        {}
                        & {} & {} &
                                {} & \ve{c} \ar@{|>}[uu]|{AND} & {} \\
\\
                \ve{a} \ar@{|>}[uu] & {} & \ve{b} 
                \ar@{|>}@/_0.3pc/[uuuull] &
                        {} & {} & {} &
                                \ve{a} \ar@{|>}[uur] & {} \ar@{|>}@{}[uu]|{OR} & \ve{b}
\ar@{|>}[uul] \ar@{|>}@/_1pc/[uuuul]
}
$$
}\end{example}
\begin{example}{\rm  \label{ex:nonedc}
It is also possible to have 
extremal realisations in which an event depends on 
an event of the family having been enabled in two distinct ways, as in the following extremal realisation with top on the left.  
$$\small
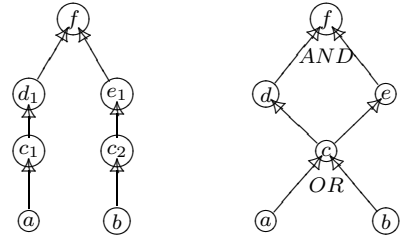
\xymatrix@R=2pt@C=4pt{
{} & \ve{f} & {} &
        {} & {} & {} & {} &
                {} & \ve{f} & {} \\
{} & 
 & {} &
        {} & {} & {} & {} &
                {} & AND & {} \\
\ve{d_1} \ar@{|>}[uur] & {} & \ve{e_1} \ar@{|>}[uul] &
        {} & {} & {} & {} &
                \ve{d} \ar@{|>}[uur] & {} & \ve{e} \ar@{|>}[uul] \\
{} & {} & {} &
        {} 
        & {} & {} & {} &
                {} & {} & {} \\
\ve{c_1} \ar@{|>}[uu] 
& {} & \ve{c_2} \ar@{|>}[uu] &
        {} & {} & {} & {} &
                {} & \ve{c} \ar@{|>}[uur] \ar@{|>}[uul] & {} \\
{} & {} & {} &
        {} & {} & {} & {} &
                {} & OR & {} \\
\ve{a} \ar@{|>}[uu] & {} & \ve{b} \ar@{|>}[uu] &
        {} & {} & {} & {} &
                \ve{a} \ar@{|>}[uur] & {} & \ve{b} \ar@{|>}[uul]
}$$
The extremal describes the event $f$ being enabled by $d$ and $e$ where they are in turn enabled by different ways of enabling $c$.  (Such phenomena will be disallowed in \edc's.)
}\end{example}
\subsection{A right adjoint to $\ges$}
The right adjoint $\er:\GES \to \ESE$ is defined on objects as follows.  
Let $A$ be a general event structure.  
Define $\er(A) = (P,\Con_P, \leq_P, \equiv_P)$ where
\begin{itemize}
\item
$P$ consists of a choice from within each isomorphism class of those extremals $p$ of $\iconf A$ with a top element---we write $\max_A(p)$ for the image of the top element in $A$;
\item
Causal dependency $\leq_P$ is $\preceq$ on $P$;
\item
$X\in \Con_P$ iff $X\fsubseteq P$ and $\max_A[X] \in \iconf A$ ---the set $[X]$ is the $\leq_P$-downwards closure of $X$; 
\item
$p_1\equiv_P p_2$ iff $p_1, p_2\in P$ and $\max_A(p_1) = \max_A(p_2)$.
\end{itemize}

\begin{prop} \label{prop:domianofrealzns}
The configurations of $P$, ordered by inclusion, are order-isomorphic to the order of extremal realisations of $\iconf A$: an extremal realisation $\rho$ corresponds, up to isomorphism, to the configuration $\set{p\in P}{p\preceq \rho}$ of $P$;  conversely, a configuration $x$ of $P$ corresponds to an extremal realisation
$\max_A:x  \to A$ with carrier $(x,\preceq)$, the restriction of the order of $P$ to $x$.
\end{prop}
 
From the above proposition we see that the events of $\er(A)$ correspond to completely-prime extremal realisations~\cite{NPW}. Henceforth we shall use the term `prime extremal' instead of the clumsier `extremal with top element.'

The component of the counit of the adjunction at $A$ is given by the function 
  $
  \max_A
$
 which determines a map $\max_A:\ges(\er(A)) \to A$ of general event structures.

\begin{theorem}
Let $A\in \GES$.  
For all $f: \ges(Q)\to A$ in $\GES$, there is a map $h:Q\to \er(A)$ in $\ESE$ such that 
$f = \max_A\circ \ges(h)$
\ie~so the diagram
$$
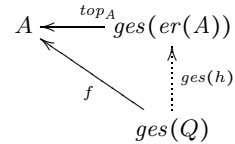
\xymatrix{
A & \ar[l]_{\max_A} \ges(\er(A))\\
& \ar[ul]^f \ges(Q) \ar@{..>}[u]_{\ges(h)}
}$$
commutes.  Moreover, if  $h':Q\to \er(A)$ is a map in $\ESE$ such that $f = \max_A\circ \ges(h')$,  then $h'\equiv h$.
\end{theorem}

The theorem does not quite exhibit a standard adjunction, because the usual cofreeness condition specifying an adjunction is weakened to only having uniqueness up to $\equiv$.  However the condition it describes does specify an exceedingly simple case of   {\em pseudo adjunction} between 2-categories---a set together with an equivalence relation is a very simple example of a category (see Appendix~A).  As a consequence, whereas with the usual cofreeness condition allows us to extend the right adjoint to arrows, so obtaining a functor, in this case following that same line will only yield a pseudo functor $\er$ as right adjoint: thus extended, $\er$ will only necessarily preserve composition and identities up to $\equiv$.

The pseudo adjunction from $\ESE$ to $\GES$ cuts down to a reflection (\ie~
the counit is a natural isomorphism) when we restrict to the subcategory of $\GES$ where all general event structures are replete.  Its right adjoint provides a pseudo functor 
embedding replete general event structures (and so families of configurations) in \ese's.

\begin{example}{\rm
On the right we show a general event structure and on its left the \ese~which it gives rise to under $\er$:
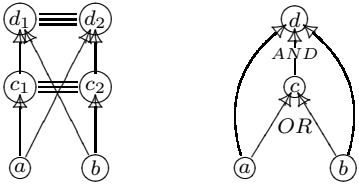
$$ \small\xymatrix@R=4pt@C=4pt{
 \ve{d_1} \ar@{|>}@3{-}[rr] & {} & \ve{d_2}
{} &  {} & {} & {} & {} &
{} & \ve{d} & 
                \\
{} & {} & {} &
        {} & {} & {} & {} &
                {} & {} & {} \\            
                \ve{c_1} \ar@{|>}[uu]^{
                } \ar@{|>}@3{-}[rr] & {} & \ve{c_2} \ar@{|>}[uu]_{
                }
& {}
& {} & {} & {} &
 {} & \ve{c} \ar@{|>}[uu]|{AND
} & {} 
                \\
{} & & {} &
        {} & {} & {} & {} &
                {} & {OR}  & {} \\
 \ve{a} \ar@{|>}[uu] \ar@{|>}[uuuurr] & {} & \ve{b} \ar@{|>}[uu] \ar@{|>}[uuuull]
& {} & {} & {} & {} &
 \ve{a} \ar@{|>}[uur] \ar@{|>}@/^1pc/[uuuur] & {} & \ve{b} \ar@{|>}[uul] \ar@{|>}@/_1pc/[uuuul]        
        }$$
}\end{example}
 
 \section{EDC'S} 
Our major motivation in developing and exploring \ese's was in order to extend strategies with parallel causes while maintaining the central operation of hiding.    What about the other 
 operation   key to the composition of strategies, \viz~pullback?  

It is well-known to be hard to construct limits such as pullback within prime event structures, so that we often rely on first carrying out the constructions in stable families.  It is sensible to seek an analogous way to construct pullbacks or pseudo pullbacks in $\ESE$.  
\subsection{Equivalence families}
In fact, the pseudo adjunction from $\ESE$ to $\GES$ factors through a more basic pseudo adjunction to families of configurations which also bear an equivalence relation on their underlying sets.  An {\em equivalence-family}  (ef) is a family of configurations $\A$ with an equivalence relation $\eeq_A$ on its underlying set $\bigcup\A$.  We can identify a family of configurations $\A$ with the 
\ef~$(\A, =)$, taking the equivalence to be simply equality on the underlying set.
 A map $f: (\A, \eeq_A)\to(\B, \eeq_B)$  between \ef's is a partial function $f:A\parrow B$ between their underlying sets which  preserves $\eeq$ 
 so that
$$x\in \A 
\,\Rightarrow\, f x\in \B\ \& \ 
\all a_1, a_2 \in x. \   f(a_1)  \eeq_B f(a_2) 
\Rightarrow a_1\eeq_A a_2\,.
$$
Composition is composition of partial functions.  We regard two maps $$f_1,  f_2: (\A, \eeq_A)\to(\B, \eeq_B)$$ as equivalent, and write $f_1\equiv f_2$,   iff they are equidefined and yield 
equivalent results. Composition respects $\equiv$.   This yields a 
  category of equivalence families $\FAME$  enriched in the category of sets with equivalence relations.  
  
  Clearly we can regard an ese $(P,
  \eeq_P)$ as an ef $(\iconf P, \eeq_P)$ and a function which is a map  of ese's as a map between the associated ef's, and this operation forms a functor.  The functor has a pseudo right adjoint built from causal realisations in a very similar manner to $\er$.  The configurations of a general event structure form an ef with the identity relation as its equivalence.  This operation is functorial and has a left adjoint which collapses an ef to a general event structure in a similar way to $\ges$; the adjunction is enriched in equivalence relations. In summary, the pseudo adjunction 
  $$
\xymatrix{
\ESE  \ar@/_/[rr]_{\ges}^{\top} &&  \GES \ar@/_/[ll]_{\er}
}
$$
 factors into a pseudo adjunction followed by an adjunction 
  $$\xymatrix{
\ESE \ar@/_/[rr]_{}^{\top} &&  \FAME \ar@/_/[ll]_{ }\ar@/_/[rr]_{}^{\top} &&   \GES  \ar@/_/[ll]_{} \,.
}$$

$\FAME$ has pullbacks and pseudo pullbacks which are easy to construct.  For example, 
 let $f: \A \to \C$ and $g:\B\to\C$ be total maps of ef's. Assume $\A$ and $\B$ have underlying sets $A$ and $B$.  Define  $D\eqdef\set{(a,b)\in A\times B}{f(a)\eeq_C g(b)}$ with projections $\pi_1$ and $\pi_2$ to the left and right components.  On $D$, take $d\eeq_D d'$ iff $\pi_1(d)\eeq_A \pi_1(d')$ and  $\pi_2(d)\eeq_B \pi_2(d')$.  Define a family of configurations of the {\em pseudo pullback} to consist of 
$x\in \D$ iff 
$ x\subseteq D$  such that
$
\pi_1 x \in \A  \ \& \  \pi_2 x \in \B\,,
$
and
$$
\eqalign{
&\all d\in x \exists d_1, \cdots, d_n\in x.\ d_n =d \ \&\ 
%
\cr
&\all i \leq n.\   \pi_1\setof{d_1,\cdots, d_i}\in \A \ \&\ \pi_2\setof{d_1,\cdots, d_i}\in\B\,.}
$$
The \ef~$\D$ with maps $\pi_1$ and  $\pi_2$ is the pseudo pullback of $f$ and $g$.  It would coincide with pullback if  $\eeq_C$ were the identity.

But unfortunately (pseudo) pullbacks in $\FAME$ don't provide us with  (pseudo) pullbacks in $\ESE$ because the right adjoint is only a pseudo functor;  in general it will only carry 
pseudo pullbacks to bipullbacks.  While $\ESE$ does have bipullbacks (in which commutations and uniqueness are only up to the equivalence $\eeq$ on maps) it doesn't always have pseudo pullbacks or pullbacks---Appendix~B.  Whereas pseudo pullbacks and pullbacks are characterised up to isomorphism, bipullbacks are only characterised up to a weaker equivalence, that induced  on objects by the equivalence on maps.
While we could develop strategies with parallel causes in the broad context of \ese's in general, doing so  would mean that the composition of strategies that ensued was not defined up to isomorphism.  This in turn would weaken our intended definition and characterisation of  such strategies as those maps into games which are stable under composition with copycat. 
 \subsection{Edc's defined}
 Fortunately there is a subcategory of $\ESE$ which supports hiding, pullbacks and pseudo pullbacks.  Define $\EDC$  to be the subcategory of $\ESE$ with objects \ese's satisfying
 $$p_1, p_2\leq p \ \&\ p_1\eeq p_2 \implies p_1= p_2\,.$$
 We call such objects {\em event structures with disjunctive causes} (\edc's).  In an \edc~an event can't causally depend on two distinct prime causes of a common disjunctive event, and so rules out realisations such as that illustrated in Example~\ref{ex:nonedc}.   
In general, within $\ESE$ we lose the local injectivity property that we're used to seeing for maps of event structures;  the maps of event structures are injective from configurations, when defined.  However for $\EDC$ we recover local injectivity w.r.t.~prime configurations:
if $f:P\to Q$ is a map in $\EDC$, then
$$
p_1,p_2 \leq_P p \ \& \ f(p_1) = f(p_2) \implies p_1 = p_2\,.
$$
The factorisation property associated with hiding in $\ESE$ is inherited by $\EDC$.

As regards (pseudo) pullbacks, we are 
fortunate in that  the complicated pseudo adjunction between \ese's and \ef's restricts down to a much simpler adjunction, in fact a coreflection, between \edc's and {\em stable} \ef's.
In an equivalence family $(\A, \eeq_A)$ say a configuration $x\in \A$ is {\em unambiguous} iff
$$
\all a_1, a_2\in x.\ a_1\eeq_A a_2 \implies a_1 = a_2\,.
$$
An equivalence family $(\A, \eeq_A)$, with underlying set of events $A$, is {\em stable} iff it satisfies
$$
\eqalign{
&\all x,y,z\in\A.\ x, y\subseteq z \ \&\ z \hbox{ is unambiguous } 
\,\Rightarrow\, x\cap y\in \A
\hbox{ and }\cr
&
\all a\in A, x\in\A.\ a\in x 
\,\Rightarrow\,  \exists z\in\A.\ z \hbox{ is unambiguous }\ \&\ a\in z\subseteq x\,.
}$$
In effect a stable equivalence family contains a stable subfamily of unambiguous configurations out of which all other configurations are obtainable as unions.  Local to any unambiguous configuration $x$ there is a partial order on its events $\leq_x$:  each   $a\in x$ determines a {\em prime configuration}
$$
[a]_x \eqdef \bigcap\set{y\in \A}{a\in y\subseteq x}\,,
$$
the minimum set of events on which $a$ depends within $x$;  taking
$a\leq_x b$ iff $[a]_x\subseteq [b]_x$ defines causal dependency between  $a, b\in x$. 
Write $\SFAME$ for the subcategory of stable \ef's.  

(Pseudo) pullbacks in stable ef's are obtained from those in \ef's simply by restricting to those configurations which are unions of unambiguous configurations.  

The configurations of an \edc~with its equivalence are easily seen to form a stable \ef~providing a full and faithful embedding of   $\EDC$ in $\SFAME$.  The embedding has a right adjoint $\Pr$.  
It is  
built out of prime extremals but we can take advantage of the fact that 
in a stable \ef~unambiguous prime extremals have the simple form of prime configurations. 
From a stable \ef $(\A, \eeq_A)$ we produce an \edc~$\Pr(\A, \eeq_A) =_{{\rm def}} (P, \Con, \leq, \eeq)$ in which  $P$ comprises the prime configurations with 
$$
\eqalign{
&[a]_x \eeq [a']_{x'} \hbox{ iff } a\eeq_A a'\ ,\cr
&
Z \in \Con  \hbox{ iff }  Z \subseteq P \ \&\ \bigcup Z \in\Fam\  \hbox{ and},\cr
&
p\leq p' \hbox{ iff }  p, p'\in P\ \&\ p\subseteq p'\,.}
$$
The adjunction is enriched in the sense that its natural bijection preserves and reflects the equivalence on maps:
$$
\xymatrix{
\EDC  \ar@/_/[rr]_{ }^{\top} &&  \SFAME \ar@/_/[ll]_{\Pr}
}
$$
  
We can now obtain a (pseudo) pullback in  \edc's 
by first forming the (pseudo) pullback of the stable \ef's obtained as their configurations
and then taking its image under the right adjoint $\Pr$.  
We now have the constructions we need to support strategies based on \edc's.

\subsection{Coreflective subcategories of 
edc's}\label{sec:subcatofedc}
 
 $\EDC$ is a coreflective subcategory of $\ESE$; the right adjoint simply cuts down to those events satisfying the \edc~property.  In turn $\EDC$ has a coreflective subcategory $\ESE^0$ comprising  those \edc's which satisfy
 $$\setof{p_1, p_2}\in\Con \ \&\ p_1\eeq p_2 \implies p_1= p_2\,.$$
 Consequently its maps are traditional maps of event structures which preserve the equivalence.
 We derive adjunctions 
 $$\xymatrix{
\ESE^0 \ar@/_/[rr]_{}^{\top} && \EDC \ar@/_/[rr]_{}^{\top}  \ar@/_/[ll]_{}&& \ESE \ar@/_/[rr]_{\ges}^{\top}  \ar@/_/[ll]_{} &&  
\GES\,.  \ar@/_/[ll]_{\er}
}
$$
 Note the last is only a pseudo adjunction.  Consequently we obtain a pseudo adjunction from $\ESE^0$, the a category of prime event structures with equivalence relations and general event structures---this makes good the promise of Section~\ref{sec:relprimeandgen}.  Inspecting the composite of the last two adjunctions, we also obtain the sense in which replete general event structures embed via a reflection in \edc's.  
   
 There is an obvious `inclusion' functor from the category of prime event structures $\PES$ to the category $\EDC$;  it extends an event structure with the identity equivalence.   Regarding $\EDC$ as a plain category, so dropping the enrichment by equivalence relations, the `inclusion' functor
$$
 \PES\hookrightarrow \EDC
$$
has a right adjoint, 
\viz~the forgetful functor which given an \edc~$P=(P,\leq,\Con, \eeq)$ produces an event structure $P_0=(P, \leq, \Con')$ by dropping the equivalence $\eeq$  
and modifying the consistency relation to 
$$X\in\Con' \hbox{ iff } X\subseteq P\ \&\ X\in\Con\ \&\ p_1\not\eeq p_2, \hbox{ for all } p_1,p_2\in X\,.$$

\noindent
The configurations of $P_0$ are the unambiguous configurations of $P$.  
 The adjunction is  a coreflection because the inclusion functor is full.
 Of course it is not   the case that the adjunction is enriched: the natural bijection of the adjunction cannot respect the equivalence on maps; 
  it cannot   compose with the pseudo adjunction from $\EDC$ to $\GES$ to yield a pseudo adjunction from $\PES$ to $\GES$.  
 
 Despite this the adjunction from $\PES$ to $\EDC$ has many useful properties.  Of importance for us is that the functor forgetting  equivalence will preserve all limits and especially pullbacks. It is helpful in relating composition of \edc-strategies to the composition of strategies based on prime event structures in~\cite{lics11}.  In composing strategies in edc's we shall only be involved with pseudo 
 pullbacks of maps $f:A\to C$ and $g:B\to C$ in which $C$ is essentially an event structure, \ie~an edc in which the equivalence is the identity relation.  The construction of such pseudo pullbacks coincides with that of pullbacks. 
  While this does not entail that composition of strategies is preserved by the forgetful functor---because the forgetful functor does not commute with hiding---it will give us a strong relationship, expressed as a map, between composition of the two kinds of strategies (based on \edc's and based on prime event structures) after and before applying the forgetful functor.   This has been extremely useful in some proofs, in importing results from~\cite{lics11}.

 \section{Strategies based on edc's}
 We develop strategies in edc's in a similar way to that of strategies in~\cite{lics11}, \viz~as certain maps stable under composition with copycat.  But 
 what is copycat on an edc? If games are edc's, shouldn't composition be based on pseudo pullback rather than pullback? To 
 separate concerns
 and, at least  initially, avoid such issues 
 we assume that games are (the edc's of) prime event structures
 , ensuring  that in 
 our uses of pullbacks they will coincide with pseudo pullbacks.  
 
 An {\em edc with polarity} comprises $(P, 
 \eeq,\pol)$, an edc~$(P, 
 \eeq)$ in which each element $p\in P$ carries a polarity $\pol(p)$ which is $+$ or $-$, according as it represents a move of Player or Opponent, and where the equivalence relation $\eeq$ respects polarity. 
   
   A {\em map} of edc's with polarity is a map of the underlying edc's which preserves polarity when 
   defined.  The adjunctions of the previous chapter are undisturbed by the addition of polarity.
   
   There are two fundamentally important operations on two-party games.  One is that of forming the dual game in which the moves of Player and Opponent are reversed.  On an \edc~with polarity $A$ this amounts to reversing the polarities of events to produce the dual $A^\perp$.  
The other operation is a simple parallel composition of games, achieved on \edc's with polarity $A$ and $B$ by simply juxtaposing them, ensuring a finite subset of events is consistent if its overlaps with the two games are individually consistent, to form $A\vvbar B$.   
  
 A game is represented by an edc with polarity in which the edc is that of a prime event structure.  A {\em pre-strategy} in \edc's, or an {\em edc pre-strategy}, in a game $A$ is a total map $\sig:S\to A$  of \edc's.   A pre-strategy from a game $A$ to a game $B$ is a pre-strategy in the game $A^\perp\vvbar B$. 
 We shall shortly refine the notion of pre-strategy  to strategy. By a strategy in a game we will mean a strategy for Player.  A strategy for Opponent,  or a counter-strategy,  in  a game $A$ will be identified with a strategy in $A^\perp$.  A map 
 $f: \sig \Rightarrow \sig'$ of edc pre-strategies $\sig:S\to A$ and $\sig':S'\to A$ is a map $f:S\to S'$ of \edc's with polarity such that  $\sig=\sig'f$;  in the standard way this determines isomorphisms of \edc~pre-strategies, important for us in a moment.

 \subsection{Copycat}
An important example of a strategy is the {\em copycat} strategy for a game $A$.  This is a strategy in the game $A^\perp\vvbar A$ which, following the spirit of a copycat, has   Player moves  copy the corresponding Opponent moves in the other component.  In more detail, the copycat strategy  comprises $\cc_A:\CC_A\to A^\perp\vvbar A$ where $\CC_A$ is obtained by 
adding extra causal dependencies to $A^\perp\vvbar A$ so that any Player move in either component  causally depends on its copy, an Opponent move, in the other~\cite{lics11}. This generates a partial order of causal dependency.  A finite set is taken to be consistent if its down-closure w.r.t.~the order generated is consistent in $A^\perp\vvbar A$; the map $\cc_A$ is the identity function on events.   We illustrate the construction on the simple game comprising a Player move causally dependent on a single Opponent move: 
$$
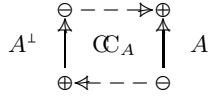
\xymatrix@R=1pt@C=1pt{
&\opmove \ar@{--|>}[rr]&& \plmove& \\\
A^\perp&&\CC_A&&A\\
&\plmove\ar@{|>}[uu] && \opmove\ar@{|>}[uu] \ar@{--|>}[ll]&
}
$$

In characterising the configurations of the copycat strategy an important partial order on configurations is revealed.  Clearly configurations of a game $A$ are ordered by inclusion $\subseteq$.  For configurations $x$ and $y$, write $x\subseteq^- y$ and $x\subseteq^+ y$ when all the additional events of the inclusion are purely Opponent, respectively,  Player moves.
A configuration $x$  of $\CC_A$ is also a configuration of $A^\perp\vvbar A$ and as such splits into two configurations  $x_1$ on the left and $x_2$ on the right. The extra causal constraints of copycat ensure that the configurations of $\CC_A$ are precisely those configurations of $A^\perp\vvbar A$ for which 
it holds that 
$$ 
 x_2\bel_A x_1 \,, \hbox{ defined as }
%
x_2 \supseteq^- x_1\cap x_2 \subseteq^+ x_1\,.
$$
Because it generalises 
the pointwise order of domain theory, initiated by Dana Scott, we have called 
$\bel_A$ the {\em Scott order}.  

 \subsection{Composing edc pre-strategies}\label{sec:comp}
  In composing two edc pre-strategies one $\sig$ in $A^\perp\vvbar B$ and another $\tau$ in $B^\perp\vvbar C$  
one firstly instantiates the Opponent moves in component $B$ by Player moves in $B^\perp$ and {\it vice versa}, and then secondly hides the resulting internal moves over $B$.  The first step is achieved efficiently via pullback.
Temporarily ignoring polarities, the pullback  in edc's
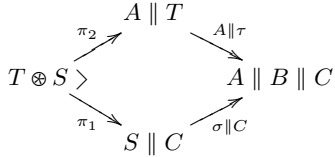
  \[
\xymatrix@R=10pt@C=10pt{
&A\parallel T	\ar[dr]^{A\parallel \tau}\\
T\sncirc S	\ \ 	\ar[ur]^{\pi_2}
		\ar[dr]_{\pi_1}
		\ar@{}[rr]|{ \ \ }
		\pb{0}&&
A\parallel B \parallel C\\
&S\parallel C	\ar[ur]_{\sigma \parallel C}
}
\]
 ``synchronises'' matching moves of $S$ and $T$ over the game $B$.  But we require a strategy over the game $A^\perp\vvbar C$ and the pullback $T\sncirc S$ has internal moves over the game $B$.  We achieve this via the projection  of  $T\sncirc S$ to its moves over $A$ and $C$.  We make use of the partial map from $A\vvbar B\vvbar C$ to $A\vvbar C$ which acts as the identity function on $A$ and $C$ and is undefined on $B$. 
The composite partial map
 
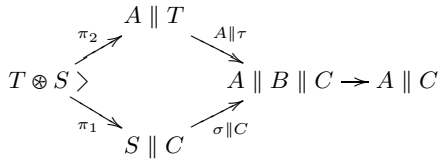
\[
\xymatrix@R=10pt@C=10pt{
&A\parallel T	\ar[dr]^{A\parallel \tau}\\
T\sncirc S\ 	\ 			\ar[ur]^{\pi_2}
		\ar[dr]_{\pi_1}
		\ar@{}[rr]|{ \ \ }
		\pb{0}&&
A\parallel B \parallel C
		\ar[r]&
A\parallel C\\
&S\parallel C	\ar[ur]_{\sigma \parallel C}
}
\]
has  defined part, yielding the composition $$\tau\scirc \sig: T\scirc S \to A^\perp\vvbar C$$ once we reinstate polarities.  The composition of edc strategies $\tau\scirc \sig$ is a form of 
synchronised composition of processes followed by the hiding of internal moves,
a view promulgated by Abramsky within traditional game semantics of programs.  

\subsection{Edc strategies}\label{sec:edcstrats}
 The article~\cite{lics11} 
 characterises through the properties of ``innocence'' and ``receptivity" those pre-strategies based on event structures which are stable under composition with the copycat strategy; 
 the characterisation 
 becomes the definition of concurrent strategy.  
 We imitate~\cite{lics11}  and provide necessary and sufficient conditions for a pre-strategy in edc's to be stable up to isomorphism under composition with copycat.  Fortunately we can inherit a great deal from the proof of~\cite{lics11} via the coreflection of event structures in \edc's of Section~\ref{sec:subcatofedc}.

An edc pre-strategy $\sig: S\to A$  is an {\em edc strategy} if it satisfies the following axioms:\\
{\em innocence:}  $ \sigma(s)\imc \sigma(s')$
if
$s\imc s'\ \&\ \pol(s) = +$ or $\pol(s') = -$\,.\\
{\em $\exists$-receptivity:} 
if $\sigma x  \longcov a$ in\, $\conf A$ with $\pol_A(a) = -$ then   
$x\longcov s  \ \&\
\\ \sig(s) = a$, for some $s\in S$\,.
(Unlike ``receptivity'' of~\cite{lics11} we do not have uniqueness.) \\
{\em +-consistency:}
$X\in \Con_S$ if $\sig X\in\Con_A$ and $[X]^+\in\Con_S$, where $X\fsubseteq S$.
(The set 
$[X]^+$ comprises the +ve elements in the downwards closure of $X$.)\\
{\em non-redundancy:} 
$s_1 = s_2$ if $[s_1)=[s_2)\ \&\ s_1\eeq_S s_2\ \&\ \pol_S(s_1)=\pol_S(s_2) 
= -$\,.\\
{\em $\eeq$-saturation:}  
$s_1 \eeq_S s_2$ if $\sig(s_1)= \sig(s_2)$\,.
   
\begin{theorem} Let $\sig:S\to A$ be an edc pre-strategy.  Then, 
$\sig \iso \cc_A\scirc \sig$ iff $\sig$ satisfies the axioms above. 
\end{theorem}

\begin{corollary}
 Let $\sig:S\to B^\perp\vvbar C$ be an edc pre-strategy. 
 Then, 
$\sig \iso \cc_C\scirc \sig\scirc \cc_B$ iff $\sig$ satisfies the axioms above. 
\end{corollary}

 
We obtain a bicategory  in which the objects are  games, the arrows $\sig:A\profto B$ are \edc~strategies $\sig$ from $A$ to $B$ and 2-cells are total maps of pre-strategies with 
  vertical composition 
  their usual composition.  Horizontal composition 
is given by composition  $\scirc$, which extends to a functor on 2-cells via the universality of pullback and the factorisation property of hiding.  
 An \edc~strategy $\sigma:A\profto B$ corresponds to its dual $\sigma^\perp:B^\perp \profto A^\perp$,   
 yielding (a bicategorical variant of) 
 compact-closure though this can weaken to $*$-autonomy with the addition of extra structure such as winning conditions or pay-off.

An edc strategy $\sig:S\to A$ is {\em deterministic}  if $S$ is deterministic as an edc with polarity: 
  $$
\all X\fsubseteq S.\  [X]^- \in \Con_S \implies X\in\Con_S\,,
$$
where
$[X]^-$ is all the Opponent moves in the down-closure $[X]$;    in other words, consistent behaviour of Opponent implies consistent behaviour. $S$ being deterministic is equivalent to $$
x\longcov{s_1} \ \& \ x\longcov{s_2} \ \&\ \pol(s_1)=+ \,\Rightarrow\, x\cup\setof{s_1, s_2} \in\conf S\,.
$$
for all $x\in\conf S, s_1, s_2\in S$.
 Copycat strategies $\cc_A$ are deterministic  iff the game $A$ is {\em 
  race-free:} 
 if  $x\longcov{a}$ and $x\longcov{a'}$ in $\conf{A}$ with $a$ and $a'$ of opposing polarities,  then
  $x\cup \{a, a'\} \in \conf{A}$. 
We obtain  a sub-bicategory of deterministic \edc~strategies between race-free games~\cite{lics11}.  
 
Such parallel deterministic strategies include the strategy sketched informally in the Introduction
in which Player makes a move iff Opponent makes one or more of their moves: \\ 
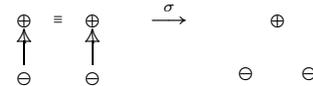
 $$
\small
\xymatrix@R=11pt@C=4pt{
\plmove
\ar@{}[rr]|\eeq&&\plmove
\\
\ar@{|>}[u] \opmove
& &\ar@{|>}[u]\opmove
}
 \qquad \arr\sig \qquad
\small
\xymatrix@R=9pt@C=0pt{
&\plmove&\\
 \opmove
 & &\opmove
 }
$$
Along the same lines there is a parallel deterministic  strategy for computing ``parallel or.''  

\section{Probabilistic edc strategies}

\subsection{Probabilistic event structures}

A probabilistic event structure essentially comprises an event structure together with a  continuous valuation on the Scott-open sets of its domain of configurations.\footnote{A {\em Scott-open} subset of configurations is upwards-closed w.r.t.~inclusion and such that if it contains the union of a directed subset $S$ of configurations then it contains an element of $S$. 
A {\em continuous valuation} is a function $w$ from  the Scott-open subsets of  $\iconf E$ to $[0,1]$ which is
 {\em (normalized)} \  $w(\iconf E) = 1$;    {\em (strict) }\  $w(\emptyset) = 0 $;
 {\em  (monotone)} \ $U \subseteq V \implies w(U)\leq w(V)$;
 \noindent{\em  (modular)} \ $w(U \cup V) + w(U\cap V) = w(U) + w(V)$; and
 \noindent {\em  (continuous) }\ $w(\bigcup_{i\in I} U_i) = {\rm sup}_{i\in I} w(U_i)$, 
 for {\em directed} unions. The idea:  $w(U)$ is the probability of a result in open set $U$.
 } The continuous valuation assigns a probability to each open set and can then be extended to a probability measure on the Borel sets~\cite{jonesplotkin}.  However open sets are several levels removed from the events of an event structure, and an equivalent but more workable definition is obtained by considering the probabilities of sub-basic open sets, generated by single finite configurations;  for each finite configuration $x$ this specifies $\Prob(x)$ the probability of obtaining events $x$, so 
 as result a
 configuration which extends the finite configuration $x$.  Such  valuations on configuration determine the continuous valuations from which they arise,  and can be characterised through the device of ``drop functions'' which measure the drop in probability across certain generalised intervals.  The characterisation yields a workable general definition of probabilistic event structure as event structures with {\em configuration-valuations}, \viz~functions from finite configurations to the unit interval for which the drop functions are always nonnegative~\cite{Probstrats
}.

In detail, a {\em probabilistic event structure} comprises an event structure $E$ with  a {\em configuration-valuation}, a function  $v$ from the finite configurations of $E$ to the unit interval  which  is 
 \begin{itemize}
\item[]
{\em (normalized)}\  $v(\emptyset) =1$   and has 
\item[] 
{\em (non\,$-$ve drop)}\ 
  $\drp y{x_1, \cdots,x_n}  \geq 0$  when $y\subseteq x_1, \cdots,x_n$ for finite configurations $y, x_1, \cdots,x_n$ of  $E$, 
\end{itemize}
where the ``drop'' across the generalized interval starting at $y$ and ending at one of the  $x_1, \cdots,x_n$ is given by 
 %
$$ 
\drp  y{x_1, \cdots,x_n}  \eqdef v(y) - \sum_I (-1)^{|I|+1} v(\bigcup_{i\in I} x_i) 
$$
---the index $I$ ranges over nonempty $I \subseteq \setof{1,\cdots, n}$ such that the union $\bigcup_{i\in I} x_i$ is a configuration. The ``drop''  $\drp  y{x_1, \cdots,x_n}$ gives the probability of the result being a configuration which includes the configuration $y$ and does not include any of the configurations $x_1, \cdots,x_n$.

If $x\subseteq y$ in $\conf E$, then  $\Prob(y\mid x) = v(y)/v(x)$; this is the probability that the resulting configuration includes the events $y$ conditional on it including the events $x$.  

\subsection{Probability with an Opponent}
This prepares the ground for a general definition of distributed probabilistic strategies, based on \edc's.    Firstly though, we should restrict to race-free games, in particular because without copycat being deterministic there would be no probabilistic identity strategies.   
A probabilistic \edc~strategy  in a game $A$,    is an \edc~strategy $\sig:S\to A$ in which we
  endow $S$ with probability, while 
 taking account of the fact that in the strategy Player can't be aware of the probabilities assigned by Opponent.   
We do this through extending the definition of  configuration-valuation via an axiom (lmc) which implies the Limited Markov Condition, LMC, of the Introduction. 
  
 Precisely,
a  {\em configuration-valuation} is now a function $v$, from finite configurations of $S$  to the unit interval, which is
 \begin{itemize}
\item[]{\em (normalized)}\  
$
 v(\emptyset ) =1$, satisfies 
\item[]
{\em (lmc)}\  
 $v(x) = v(y)$ when  $x\subseteq^- y$ for finite configurations $x$, $y$ of 
  $S$,
 and 
  the 
\item[]{\em (+ve drop condition)}\ 
 $
 \drp y{x_1, \cdots, x_n} \geq 0
 $
 when  $y\subseteq^{+} x_1, \cdots, x_n$ for finite configurations of $S$.  
  \end{itemize}
  
  When $x\subseteq^+ y$ in $\conf S$, we can still express $\Prob(y\mid x)$,  the conditional probability of Player making the moves $y\setdif x$ given $x$, as $v(y)/v(x)$.  In fact all such conditional probabilities determine $v$ via normalisation and lmc.  As $A$ is race-free it follows $S$ is also race-free. 
  Hence if $x$ is a finite configuration at which $x\longcov \plmove$ and $x\longcov\opmove$ then $x\cup\setof{\plmove,\opmove}$ is also a configuration, and both moves are $\plmove, \opmove$ are causally independent (or concurrent).  From lmc we obtain LMC directly:
 $
 \Prob(\plmove\mid x)=\Prob(x,\plmove\mid x)  = \\
 \hbox{\qquad}\qquad\qquad
 v(x\cup\setof\plmove)/v(x) =  v(x\cup\setof{\plmove,\opmove})/v(x\cup\setof\opmove) = \\  
  \hbox{\qquad}\qquad\qquad
 \Prob(x,\plmove, \opmove \mid x, \opmove) = \Prob(\plmove\mid x, \opmove)\,.
 $\\
  A dual form of LMC  will hold of a counterstrategy, a strategy for Opponent; the LMCs for Player and Opponent   will  together ensure the   
    probabilistic  independence of Player and Opponent moves from a common configuration.    
 
  A {\em probabilistic edc strategy} in race-free game $A$  comprises an edc strategy $\sig:S\to A$   with a configuration-valuation $v$ for $S$.   
 A {\em probabilistic edc strategy} between race-free games $A$ to $B$ is a probabilistic edc strategy in $A^\perp\vvbar B$.   Note that the configuration-valuation of an edc  doesn't necessarily respect the equivalence of the edc; different prime causes of a common disjunctive event may well be associated with different probabilities.

\begin{example}{\rm  Recall the game of the Introduction.  In the edc strategy \, \, 
$
w1\small
\xymatrix@R=11pt@C=4pt{
\plmove
\ar@{}[rr]|\eeq&&\plmove
\\
\ar@{|>}[u] \opmove
& &\ar@{|>}[u]\opmove
}
w2$
\, \, of Section~\ref{sec:edcstrats}
individual success of 
the two watchers may be associated with probabilities $p_1\in [0,1]$ and $p_2\in [0,1]$, respectively, and their joint success with $q\in [0,1]$ provided they form a configuration valuation $v$.  
In other words, 
$v(x) = p_1$ if $x$ contains $w1$ and not $w2$;
$v(x) = p_2$ if $x$ contains $w2$ and not $w1$;  and 
$v(x) =q$ if $x$ contains both $w1$ and $w2$;   $v(x)=1$ otherwise;   
and $p_1 + p_2 - q \leq 1$, in order to satisfy the +-drop condition.  
}
\endex\end{example}

We extend the usual composition of edc strategies to probabilistic edc strategies.  Assume probabilistic edc strategies $\sig: S\to A^\perp\vvbar B$, with configuration-valuation $v_S
$, and $\tau:T\to B^\perp\vvbar C$ with 
$v_T
$.  Their composition is defined to be $\tau\scirc \sig:T\scirc S \to A^\perp\vvbar C$ with a  configuration-valuation  $v$ given by 
$$
v(x) = v_S(\pi^S_1 x) .
v_T(\pi^T_2 x)  
$$
for $x$ a finite configuration of $ T \scirc  S $.  The configuration $\pi_1^S x$  is the  component in $\conf S$
of the projection $\pi_1 x\in \conf{S\vvbar C}$ from the pullback defined in Section~\ref{sec:comp}; similarly $\pi^T_2 x$ is the $T$-component of $\pi_2 x$. 
The proof that $v$ is indeed a configuration-valuation is quite subtle and relies heavily on properties of ``drop'' functions.
\subsection{A bicategory of probabilistic edc strategies}
We obtain a bicategory of probabilistic edc strategies in which objects are
race-free games. Maps are probabilistic edc strategies.  
 Identities are given by copycat strategies, which for race-free games are deterministic, so permit configuration-valuations which are  constantly 1.  
Generally, a probabilistic edc strategy  is {\em deterministic} if its configuration-valuation assigns 1 to all finite configurations;  its underlying edc strategy is then necessarily deterministic too. 


The 2-cells of the bicategory require consideration.  Whereas we can always ``push forward'' a probability measure from the domain to the codomain of a measurable function this is not true  generally  for configuration-valuations involving Opponent moves.  However
: 
\begin{theorem} \label{thm:pushforward}  Let $f:\sig\Rightarrow \sig'$ be a 2-cell between edc strategies $\sig:S\to A$ and $\sig':S'\to A$ which is a rigid map of event structures. 
Let $v$ be a configuration-valuation on $S$.  Taking
$
v'(y) \eqdef \sum_{x : 
f x=y} v(x)
$
for $y\in\conf{S'}$, defines a configuration-valuation, written $f v$,
 on $S'$. 
 \end{theorem}
A 2-cell from $\sig,v$ to  $\sig',v'$
 is a 2-cell $f:\sig\Rightarrow \sig'$ of edc strategies in which $f:S\to S'$ is a rigid map of event structures and for which the ``push-forward''
 $f v$ satisfies $
 (f v)(x') \leq v'(x')\,,$
 for all configurations $x'\in\conf{S'}$.  
Rigid 2-cells include rigid embeddings 
giving the machinery to define probabilistic strategies recursively.  
\section{Constructions on probabilistic edc strategies
}
Following~\cite{lics2014-2,ictac2015}, race-free games play the role of types and support operations  of forming the dual $A^\perp$, 
simple parallel composition $A\vvbar B$,
sum $\Sigma_{i\in I} A_i$ and recursively-defined games.
Terms 
have typings \\
\hbox{\ }\qquad 
$
x_1:A_1, \cdots, x_m:A_m \vdash\  t \ \dashv y_1:B_1, \cdots, y_n:B_n\ ,
$\\

\noindent
where all the variables are distinct,
and denote probabilistic \edc~strategies from the game $
 \vec A = A_1 \vvbar \cdots \vvbar A_m$   to the game $\vec B = B_1 \vvbar \cdots \vvbar B_n$.
 We can think of the term $t$ as a box with input wires $x_1,\cdots,x_m$ and output wires  $y_1, \cdots, y_n$.  The term $t$ denotes a probabilistic edc strategy $S\to \vec A^\perp \vvbar \vec B$ with configuration valuation $v$ and describes witnesses, finite configurations of $S$,   to a relation between finite configurations $\vec x$ of $\vec A$ and $\vec y$ of $\vec B$, 
together with their conditional probabilities.
The following constructions, first described for  (probabilistic) concurrent strategies in~\cite{lics2014-2,ictac2015}, extend to  (probabilistic) edc strategies, though note that duplication now becomes deterministic as an edc strategy for a broader class of games.

\noindent
{\bf Composition} $\Gamma \vdash 
\stcomp\Delta t u
 \dashv \Eta$
 if
 $\Gamma \vdash t \dashv \Delta$ and $\Delta \vdash u \dashv \Eta$.

\noindent
{\bf Probabilistic sum}
 $\Gamma \vdash \Sigma_{i \in I}  p_i  t_i \dashv  \Delta$
 if
 $\Gamma \vdash t_i \dashv  \Delta$ 
 for $i\in I$, assumed countable, and a sub-probability distribution $p_i, i\in I$.  The empty sum denotes $\bot$, the minimum strategy in the game $\Gamma^\perp\vvbar \Delta$.  
 
\noindent
 {\bf Conjunction}
  $\Gamma \vdash t_1\wedge t_2 \dashv  \Delta$ is given by pullback
  of 
  $\Gamma \vdash t_1 \dashv  \Delta$  and $\Gamma \vdash t_2 \dashv  \Delta$ from the game $\Gamma^\perp\vvbar \Delta$.
  
\noindent {\bf  Copycat 
terms} of the form $
\vec x:\vec A \vdash\  g \vec y\bel_C f \vec x\  \dashv \vec y:\vec B\,, 
$
where $f:\vec A\to C$ and $g:\vec B\to C$ are (affine) maps of event structures preserving polarity.  Such terms introduce new ``causal wiring'' and subsume copycat, injections and projections associated with sums, and prefix operations 
and can achieve the effect of $\lambda$-abstraction
on strategies~\cite{lics2014-2}.  With composition they allow us to express a {\em trace operation}.They denote deterministic edc strategies---so a probabilistic edc strategy with configuration-valuation constantly one---provided 
 $f$ reflects $-$-compatibility and 
 $g$ reflects $+$-compatibility.  The map $g$ reflects $+$-compatibility if whenever 
 $x\subseteq^+ x_1$ and  $x\subseteq^+  x_2$ in the configurations of $\vec B$
 and
 $f x_1 \cup f x_2$ is 
 a configuration,  
 then so is 
 $x_1\cup x_2$.  
Reflecting $-$-compatibility is 
analogous
.   

\noindent
{\bf Duplication}  Duplications of arguments is essential if we are to support the recursive definition of strategies.  
We  duplicate arguments through an 
\edc~strategy 
$\delta_A: A\profto A\vvbar A$.  Intuitively it behaves like  the copycat strategy but where a Player move in the left component may be caused in parallel by  either of its corresponding Opponent moves from the two components on the right.  We show $\delta_A$ when  $A$ consists of a single Player move $\plmove$ and, respectively,   a single Opponent move $\opmove$:
  
\quad\ 
$A=\plmove$,
$\xymatrix@R=0pt@C=12pt{&\plmove
\\
\opmove \ar@{|>}[ur]\ar@{|>}[dr]&\\
&\plmove
}
$
\qquad\qquad\quad
$A=\opmove$, \ 
 $\xymatrix@R=10pt@C=12pt{
\plmove\ar@{}| {{\rotatebox[origin=c]{90}{  $ \eeq$  }}}[d]&\ar@{|>}[l]\opmove\\
\plmove&\ar@{|>}[l]\opmove}
$

\noindent
The general definition is in Appendix~C. In general, duplication $\delta_A$ is deterministic iff $A$ is {\em deterministic for Opponent}, \ie~$A^\perp$ is deterministic as an edc with polarity.  Then
$\delta_A$ extends directly to a probabilistic edc strategy and is a comonoid. 
(When the duplication strategy is based on prime event structures, the duplication strategy is not  deterministic unless the game consists purely of Player moves, making
  associativity fail with the introduction of probability~\cite{ictac2015}.)  

\noindent
{\bf Recursion} Once we have duplication strategies we can treat recursion using standard machinery~\cite{icalp82}; recall that 2-cells, the maps between probabilistic strategies, include rigid embeddings, 
  so an approximation order $\tri$ of rigid inclusions. 
  The order forms a `large complete partial order' with a bottom element the minimum strategy $\bot$.   
Given $x:A, \Ga \vdash t   \dashv y:A$, 
the term $\Ga \vdash  \mu\, x\!:\!A.\, t   \dashv y:A$ denotes the
 $\tri$-least fixed point amongst probabilistic strategies $X$ in $\Ga^\perp\vvbar A$ of the $\tri$-
 continuous operation $F(X)= t\scirc (\id_\Ga\vvbar X)\scirc \delta_\Ga$.  
This requires the games $\Gamma$ are  deterministic for Opponent.

 \subsection{Special cases and extensions}  
 The constructions yield 
 {\em deterministic} edc strategies  if we avoid probabilistic sums.   
  
 If we drop probability, we can 
drop race-freeness on games, the determinacy conditions on copycat terms and parameters of recursions, and replace probabilistic  by nondeterministic  sum, to obtain constructions for {\em nondeterministic} edc strategies. 
 
Even without probability, we obtain an interesting bicategory if we restrict to games in which all moves are those of Player.
Duplication is now expressible in event structures. If we further restrict to strategies described with event structures (so the concurrent strategies of~\cite{lics11}) we obtain a monoidal-closed bicategory with simple parallel composition as tensor. With the addition of probability 
we obtain a framework for probabilistic dataflow. 
Note that probability distributions on \eg~domains of infinite streams induced by configuration-valuations can well be continuous on their maximal elements.    
Given all this the much richer types and language of the previous section 
should support a useful style of probabilistic programming 
based on probabilistic strategies.


In general, 
games can be extended to games with {\em imperfect information} and {\em pay-off} as in~\cite{Probstrats}; then they, and the probabilistic concurrent games of~\cite{Probstrats},  include Blackwell games~\cite{det-blackwell}.

 There is an alternative method for introducing {\em parallel causes via symmetry}, through a pseudo monad $?$ on games
; an edc strategy in a game $A$ corresponds to a strategy in $?A$; the monad $?$ introduces multiple symmetric parallel causes to Player moves
~\cite{ecsym-notes,CCW}.

\acks
Thanks to  Simon Castellan, Pierre Clairambault, Mai Gehrke, Jonathan Hayman and Martin Hyland for advice and encouragement, to ENS Paris for supporting Marc de Visme's internship and to the ERC for Advanced Grant ECSYM.


\bibliographystyle{abbrvnat}
\softraggedright
\bibliography{biblio}

\appendix

\section{{\bf Equiv}-enriched categories}
Here we explain in more detail what we mean when we say ``enriched in the category of of sets with equivalence relations'' and employ terms such as ``enriched adjunction,'' ``pseudo adjunction'' and ``pseudo pullback."  

$\Equiv$ is the category of {\em equivalence relations}.  Its objects are $(A,\eeq_A)$ comprising a set $A$ on which there is an equivalence relation $\eeq_A$.  Its maps $f:(A, \eeq_A)\to (B,\eeq_B)$ are total functions $f:A\to B$ which preserve equivalence.

We shall use some basic notions from enriched category theory~\cite{kelly}.  We shall be concerned with categories enriched in $\Equiv$, called  $\Equiv$-enriched categories, in which the homsets possess the structure of equivalence relations, respected by composition. 
This is the sense in which we say categories are enriched in (the category of) equivalence relations. 
We similarly borrow the concept of 
an $\Equiv$-enriched functor between $\Equiv$-enriched categories which preserve equivalence in acting on homsets.   An $\Equiv$-enriched adjunction  is a usual adjunction in which  the natural bijection preserves and reflects equivalence.

Because an object in $\Equiv$ can be regarded as a (very simple) category, 
we can regard  $\Equiv$-enriched categories as a (very simple) 2-categories to which notions from 2-categories apply~\cite{Power2-cats}.  

A {\em pseudo functor} between $\Equiv$-enriched categories is like a functor but the usual laws only need hold up to equivalence.
A {\em pseudo adjunction} (or biadjunction) between 2-categories permits a weakening of the usual natural isomorphism between homsets, now also categories,  to a natural equivalence of categories. 
In the special case of a pseudo adjunction
between $\Equiv$-enriched categories the equivalence of homset categories amounts to a pair of $\eeq$-preserving functions  whose compositions are $\eeq$-equivalent to the identity function.
With traditional adjunctions by specifying the action of one adjoint solely on objects we determine it as a functor; with pseudo adjunctions we can only determine it as a pseudo functor---in general a pseudo adjunction relates two pseudo functors. 
Pseudo adjunctions compose in the expected way. 
An $\Equiv$-enriched adjunction is a special case of a 2-adjunction between 2-categories and 
a very special case of pseudo adjunction. In this article there are many cases in which we compose an $\Equiv$-enriched adjunction with a pseudo adjunction to obtain a new pseudo adjunction.  

Similarly we can specialise the notions pseudo pullbacks and   bipullbacks from 2-categories  to $\Equiv$-enriched categories.  Let $f: A\to C$  and $g:B\to C$ be two maps in an $\Equiv$-enriched category.  A {\em pseudo pullback} of $f$ and $g$ is an object $D$ and maps
$p:D\to A$ and $q:D\to B$ such that
$f\circ  p \eeq g\circ  q $ which satisfy the further property that 
for any  $D'$ and maps $p':D'\to A$ and $q':D'\to B$ such that $f\circ  p' \eeq g\circ  q' $, there is a unique map $h:D'\to D$ such that $p' = p\circ  h$ and $q' = q\circ  h$. There is an obvious weakening of pseudo pullbacks to the situation in which the uniqueness is replaced by uniqueness up to $\eeq$ and the equalities by $\eeq$---these are simple special cases of bilimits called  {\em bipullbacks}.

Right adjoints in a 2-adjunction preserve pseudo pullbacks whereas right adjoints in a pseudo adjunction are only assured to preserve bipullbacks.  

\section{On (pseudo) pullbacks of ese's}
We show that the enriched category of \ese's $\ESE$ does not always have pullbacks and pseudo pullbacks of maps $f:A\to C$ and $g:B\to C$,  the reason why we use the subcategory $\EDC$, which does, as a foundation on which to develop strategies with parallel causes.  It suffices to exhibit the lack of pullbacks when $C$ is an (ese of an) event structure as then pullbacks and pseudo pullbacks coincide.  
Take $C$ to be 
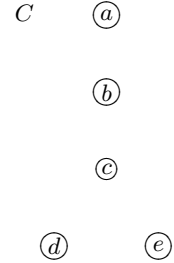
\begin{displaymath} C\  \xymatrix@C=1em@R=2em{
{} & \ve{a} & {} \\
{} & \ve{b} & {} \\
{} & \ve{c} & {} \\
\ve{d} & {} & \ve{e}
}\end{displaymath}
with $A$ and $B$ being respectively
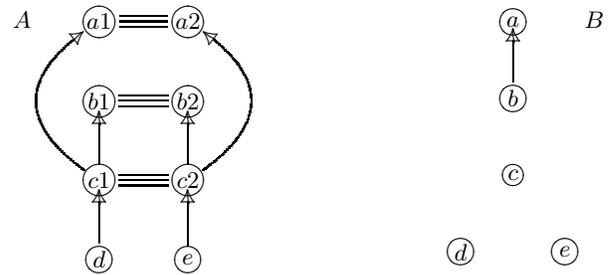
\begin{displaymath} 
A \qquad
\xymatrix@C=2em@R=2em{
\ve{a1} \ar@3{-}[r] & \ve{a2} \\
\ve{b1} \ar@3{-}[r] & \ve{b2} \\
\ve{c1} \ar@3{-}[r] \ar@{|>}[u] \ar@{|>}@/^2pc/[uu] & \ve{c2} \ar@{|>}[u] \ar@{|>}@/_2pc/[uu] \\
\ve{d} \ar@{|>}[u] & \ve{e} \ar@{|>}[u]}
\qquad\qquad\qquad\qquad\qquad
 \xymatrix@C=1em@R=2em{
{} & \ve{a} & {} \\
{} & \ve{b} \ar@{|>}[u] & {} \\
{} & \ve{c} & {} \\
\ve{d} & {} & \ve{e}
}
\ B
\end{displaymath}
with the  obvious maps $f:A\to C$ and $g:B\to C$ (given by the lettering).  In fact, $A$ and $B$ are edc's.  

The pullback in \edc's $\EDC$ does exist and is given by 
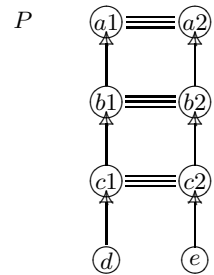
\begin{displaymath} 
P\qquad\  \xymatrix@C=2em@R=2em{
\ve{a1} \ar@3{-}[r] & \ve{a2} \\
\ve{b1} \ar@3{-}[r] \ar@{|>}[u] & \ve{b2} \ar@{|>}[u] \\
\ve{c1} \ar@3{-}[r] \ar@{|>}[u] & \ve{c2} \ar@{|>}[u] \\
\ve{d} \ar@{|>}[u] & \ve{e} \ar@{|>}[u]
}\end{displaymath}
with the obvious projection maps.  
However this is not a pullback in $\ESE$.  Consider the \ese 
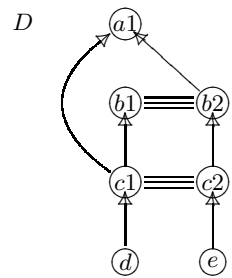
\begin{displaymath} 
D\qquad \quad 
\xymatrix@C=2em@R=2em{
\ve{a1} & {} \\
\ve{b1} \ar@3{-}[r] & \ve{b2} \ar@{|>}[lu] \\
\ve{c1} \ar@3{-}[r] \ar@{|>}@/^2pc/[uu] \ar@{|>}[u] & \ve{c2} \ar@{|>}[u] \\
\ve{d} \ar@{|>}[u] & \ve{e} \ar@{|>}[u]
}\end{displaymath}
with the obvious total maps to $A$ and $B$; they  form a commuting square with $f$ and $g$. This cannot factor through $P$:   event   $b2$ has to be mapped
to $b2$ in $P$, but then $a1$ cannot be mapped to $a1$ (it wouldn't yield  a
map) nor to $a2$ (it would violate commutation required  of a pullback). 

There is a bipullback got by applying the pseudo functor $\er$ to the pullback in \ef's:
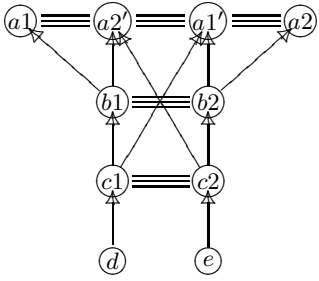
\begin{displaymath} \xymatrix@=2em{
\ve{a1} \ar@3{-}[r] & \ve{a2'} \ar@3{-}[r] & \ve{a1'} \ar@3{-}[r] & \ve{a2} \\
{} & \ve{b1} \ar@3{-}[r] \ar@{|>}[ul] \ar@{|>}[u] & \ve{b2} \ar@{|>}[u] \ar@{|>}[ur] & {} \\
{} & \ve{c1} \ar@3{-}[r] \ar@{|>}[u] \ar@{|>}[uur] & \ve{c2} \ar@{|>}[u] \ar@{|>}[uul] & {} \\
{} & \ve{d} \ar@{|>}[u] & \ve{e} \ar@{|>}[u] & {}
}\end{displaymath}
But this is not a pullback because  in the \ese~$E$ below the required mediating map is not 
  unique in that  $a1$ can go to either   $a1$ or
$a1'$:
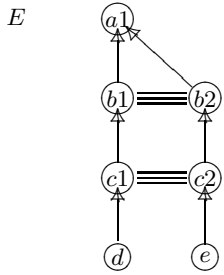
\begin{displaymath} 
E\qquad\quad  
\xymatrix@C=2em@R=2em{
\ve{a1} & {} \\
\ve{b1} \ar@3{-}[r] \ar@{|>}[u] & \ve{b2} \ar@{|>}[ul] \\
\ve{c1} \ar@3{-}[r] \ar@{|>}[u] & \ve{c2} \ar@{|>}[u] \\
\ve{d} \ar@{|>}[u] & \ve{e} \ar@{|>}[u]
}\end{displaymath}

In fact, there is no pullback of $f$ and $g$.  To show this we use an additional \ese:
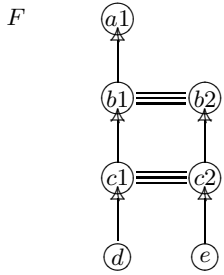
\begin{displaymath}
F \qquad\quad
 \xymatrix@C=2em@R=2em{
\ve{a1} & {} \\
\ve{b1} \ar@3{-}[r] \ar@{|>}[u] & \ve{b2}\\
\ve{c1} \ar@3{-}[r] \ar@{|>}[u] & \ve{c2} \ar@{|>}[u] \\
\ve{d} \ar@{|>}[u] & \ve{e} \ar@{|>}[u]
}\end{displaymath}

Suppose $Q$ with projection maps to $A$ and $B$ 
were a pullback of $f$ and $g$ in $\ESE$. Consider the three \ese's
 $D$, $E$ and $F$ with their obvious maps to $A$ and $B$; in each case they form a commuting square with $f$ and $g$. There are three unique maps $h_D : D \to Q$,
$h_E : E \to Q$, and $h_F : F \to Q$ such that the corresponding pullback
diagrams commute.
We remark that there are also   obvious maps $k_D : E \to D$ and   $k_F : E \to
F$ (given by the lettering) which commute with the maps to the components $A$ and $B$. By uniqueness, we have $h_D \circ k_D
= h_E = h_F \circ k_F$, so we have $h_D(a1) = h_F(a1)$.
From the definition of the maps, the event $h_D(a1) = h_F(a1)$ has at most one
$\leq$-predecessor in $Q$  which is sent to $b$ in $C$ (as $D$ only has one). Because of the projection to $B$, it has at least one (as $B$ has one). So the event $h_D(a1) = h_F(a1)$ has
exactly one predecessor which is sent to $b$. From the definition
of maps, this event is $h_D(b2)$ which equals $h_F(b1)$. But $h_D(b2)$ cannot equal $h_F(b1)$ as they go to two different events of $A$ ---a
contradiction.
Hence there can be no pullback of $f$ and $g$ in $\ESE$.
(By adding   intermediary events, we 
would encounter 
essentially the same example in the composition, before hiding, of strategies if they were to be developed within the broader category of \ese's.)

\section{The edc duplication strategy}
We present the general definition of the edc {\em duplication} strategy $\delta_A: A\profto A\vvbar A$
for a race-free game $A$.  
 
For each triple $(x, y_1, y_2)$, where 
$x\in\conf{A^\perp}$ and  
$y_1,y_2\in\conf A$,
which is {\em balanced}, \ie{} 
 $$
 \eqalign{
 &\all a \in y_1 \cup y_2.\ \pol_A(a)= + \implies   a \in x \hbox{ and }\cr
 &
 \all a\in x.\ \pol_{A^\perp}(a) = + \implies   a \in y_1 \hbox{ or }   a \in y_2\,,}
$$
and {\em choice} function 
$
\chi: x^+ \to \setof{1,2}\,,
$
from the positive events of $x$ denoted by $x^+$,
such that
$
\chi(a) = 1 \implies 
a \in y_1$ and $
\chi(a) = 2 \implies 
a \in y_2
$,
the order $q(x, y_1, y_2; \chi)$ is defined to have underlying set
$
\setof 0 \times x \,\cup\, \setof 1 \times y_1 \,\cup\, \setof 2\times y_2
$
with order generated by that inherited from $A^\perp\vvbar A\vvbar A$ together with
$$
\eqalign{
&\set{((0,
a),\, (1,a))}{a\in y_1\ \&\ \pol_A(a)=+} \ \cup\quad\cr
& \set{((0,
a), (2,a))}{a\in y_2\ \&\ \pol_A(a)=+}\ \cup\cr
&\set{((\chi(a), 
a), \, (0,a))}{a\in x \ \&\ \pol_{A^\perp}(a) = +}\,.}
$$

Now we can define $\delta_A:D_A \to A^\perp\vvbar A\vvbar A$. 
The \edc~$D_A$ comprises $(D_A, \leq, \Con, \eeq, \pol)$ with 
\begin{itemize}
\item[]
{\em events} $D_A$  consisting of all  $d=q(x, y_1, y_2; \chi)$, for balanced $(x, y_1, y_2)$ and choice function $\chi$, which have a top element $\delta_A(d)$; 
\item[]
{\em causal dependency} $d\leq d'$ iff there is a rigid inclusion map from $d$ into $d'$ (regarded as event structures);
\item[]
{\em consistency} $X\in \Con$ 
iff $X\fsubseteq D_A$ and the image of its $\leq$-down-closure, $\delta_A [X]$, is consistent in $A^\perp\vvbar A\vvbar A$;
\item[]
{\em equivalence} $d\eeq d'$ 
iff
$\delta_A(d)=\delta_A(d')$, \ie~they have the same top element in $A^\perp\vvbar A\vvbar A$; and 
\item[]
with  the {\em polarity} of events $D_A$ inherited from the polarity of their top elements, \ie~$\pol(d) = \pol_A(\delta_A(d))$ for $d\in D_A$.  
\end{itemize}

We obtain  an  edc   strategy $\delta_A: A\profto A\vvbar A$ in which 
$
\delta_A: D_A \to A^\perp\vvbar A\vvbar A
$
sends a prime to its top element.  The \edc~strategy  $\delta_A$ forms a comonoid with counit  $\bot:A\profto \emptygame$. 

The duplication strategy $\delta_A$ is deterministic iff no Opponent moves in $A$ are in immediate conflict, \ie~if
$x\longcov{a_1}$ and $x\longcov{a_2}$ in $\conf A$ and $\pol_A(a_1) = \pol_A(a_2) = -$ then $x\cup\setof{a_1, a_2}\in\conf A$.  Given that $A$ is race-free, $\delta_A$ is deterministic iff $A^\perp$ is deterministic as an \edc~with polarity---a condition we call {\em deterministic for Opponent}.  Under the condition that $A^\perp$ is deterministic, as $\delta_A$ is a deterministic edc strategy it extends directly to a probabilistic edc strategy with configuration-valuation having constant value 1.  Then the probabilstic \edc~strategy 
 $\delta_A$ forms a comonoid with counit  $\bot:A\profto \emptygame$.

\end{document}